\documentclass[10pt,aps,prl,reprint,showkeys, twocolumn,
 amsmath,amssymb,
floatfix,
]{revtex4-2}
\PassOptionsToPackage{backend=biber}{biblatex}
\usepackage[T1]{fontenc}
\usepackage[utf8]{inputenc}
\usepackage{xcolor}
\definecolor{amethyst}{rgb}{0.6, 0.4, 0.8}

\usepackage{siunitx}
\usepackage{amsmath} 
\usepackage{bm}
\newcommand{\angstrom}{\textup{\AA}}
\usepackage{hyperref}
\hypersetup{colorlinks=true, citecolor=blue, urlcolor=blue, linkcolor=black}

\usepackage{tikz}
\usetikzlibrary{shapes,arrows}

\tikzstyle{io} = [trapezium, trapezium left angle=70, trapezium right angle=110, minimum height=0.6cm, text badly centered, draw, fill=red!20]
\tikzstyle{decision} = [diamond, draw, fill=blue!20, text width=4.5em, aspect=2, text badly centered, node distance=3cm, inner sep=0pt]
\tikzstyle{block} = [rectangle, draw, fill=green!20, text width=5em, text centered, rounded corners, minimum height=4em]
\tikzstyle{section} = [rectangle, draw, fill=red!40, text width=5em, text width=8em, text centered, rounded corners, minimum height=4em, minimum width=6em]
\tikzstyle{line} = [draw, -latex']
\tikzstyle{cloud} = [draw, ellipse,fill=red!20, text centered]

\usepackage{graphicx} 
\graphicspath{ {DGraphics/} }
\usepackage{caption}
\usepackage{subfig}
\usepackage{wrapfig}

\usepackage{relsize}
\usepackage{multirow}
\usepackage{dcolumn}
\usepackage{amsthm}

\newcommand{\figref}[2][Fig.~]{#1\ref{#2}}

\renewcommand{\eqref}[2][Equation~]{#1(\ref{#2})}
 \usepackage{textcomp}

\begin{document}

\title{Augmentation of Universal Potentials for Broad Applications}

\author{Joe Pitfield}
\email{joepitfield@gmail.com}
\author{Florian Brix}

\author{Zeyuan Tang}
\author{Andreas Møller Slavensky}

\author{Nikolaj Rønne}
\author{Mads-Peter Verner Christiansen}
\author{Bjørk Hammer}
\email{hammer@phys.au.dk}

\affiliation{Center for Interstellar Catalysis, Department of Physics and Astronomy, Aarhus University, DK‐8000 Aarhus C, Denmark}
\begin{abstract}
Universal potentials open the door for DFT level calculations at a 
fraction of their cost. We find that for application to 
systems outside the scope of its training data, CHGNet\cite{deng2023chgnet} has the potential
to succeed out of the box, but can also fail significantly in predicting the ground state 
configuration. We demonstrate that via fine-tuning or a $\Delta$-learning
approach it is possible to augment the overall performance of 
universal potentials for specific cluster and surface systems. We
utilize this to investigate and explain experimentally observed defects
in the Ag(111)-O surface reconstruction and explain the mechanics behind its formation.

\end{abstract}
\keywords{Global Optimisation, Machine Learning, Neural Network Potentials, Gaussian Process Regression, Surface Reconstruction, Cluster}

\maketitle


Atomic structure plays a crucial role in the understanding and development of new 
materials and technologies. A great many different compounds have structures which 
have been determined by experimental observation through a variety of techniques
\cite{fewster2004advances,stobinski2014graphene,vcerny2017crystal,gloystein2020nanopyramidal,shields2024stm}. 
However, the configuration space of potential structures increases exponentially
 with degrees of freedom as we combine surfaces with adsorbates, 
 defects, alloys and interfaces. 
 Theoretical structure prediction offers the opportunity to screen and 
 predict the structural properties of such systems. 
 Studies of surface structure \cite{gutierrez2001theoretical,zhang2018atomistic}, 
 catalysis \cite{hendriksen2010role,polo2019surface,chen2019surface,Baker2022} and 
 interface physics \cite{hill2000nickel,schusteritsch2015first,TAYLOR2020107515,pitfield2024predicting} 
 provide valuable insight into not only the interesting material properties of these systems, 
 but also the key limitations of first principles methods in 
 addressing these ever more configurationally complex problems. 

Machine learning models offer the ability to limit the amount of expensive 
first principles calculations performed, and the possibility of extending the 
results of small calculations to larger systems without the corresponding 
cubic\cite{mohr2015accurate} increase in cost common to density functional theory.
 Machine learning has been applied in a variety of ways to alleviate the computational 
 workload of materials science: neural network potentials 
\cite{behler2007generalized,schutt2018schnet,unke2019physnet,gasteiger2021gemnet,schutt2021equivariant,batatia2022mace,batzner20223} 
 and Gaussian process regression \cite{denzel2018gaussian,koistinen2019minimum,koistinen2019nudged,deringer2021gaussian, bisbo2022global,merte2022structure,hamamoto2023machine} 
 for energetic evaluation, reinforcement\cite{jorgensen2019atomistic,zhou2019optimization,gogineni2020torsionnet} and active learning\cite{smith2018less,pitfield2024predicting} 
 for candidate structure suggestion.

Such methods reduce the demand for expensive first principles calculations 
to be performed where the chosen model is well informed. If the model is not 
well informed, such as during the initialisation of such approaches, acquiring 
sufficient data often remains expensive. Recent interest has 
surrounded the utilisation of comprehensive structural databases such as 
the Materials Project\cite{persson2012prediction}, the OpenCatalyst Project\cite{chanussot2021open} 
and PubChem\cite{kim2016pubchem} in training potentials which are broadly applicable. 
These so-called \textit{universal potentials} theoretically remove 
the expensive initialisation step of any materials science workflow, 
promising orders of magnitude decrease in cost for DFT accuracy calculations. 
Recent examples include M3GNet\cite{chen2022universal}, MACE\cite{batatia2023foundation}, 
AisNet\cite{hu2023aisnet}, CHGNet\cite{deng2023chgnet} and DPA2 \cite{zhang2023dpa}. 
This work will focus on CHGNet, as it suggests strong performance with 
relatively few parameters ($\sim$450,000).

\par{}CHGNet is trained on the Materials Project dataset, which is
dominated by structures which fall into the bulk
regime. Unsurprisingly, CHGNet has been demonstrated to perform well
when applied to materials in this realm\cite{deng2023chgnet}. However,
the majority of device physics occurs in regimes where the materials deviate from perfect
bulk crystals: interfaces have drastically varying conductivities 
\cite{marks1990interfaces,chen1997size,nan2004interface,schusteritsch2015first,burger2016review,zhang2019carbon,liu2019influence},
defects play a key role in electronics
\cite{bryan2005doped,shklovskii2013electronic,durand2015defect}, 
surfaces are defining in heterogeneous catalysis
\cite{rosenthal2011functional,huang2019surface,zaera2021molecular}, and
nano-particulate phases are key in atmospheric\cite{biswas2005nanoparticles} 
and astro-chemical\cite{tang2022top,rasmussen2023gas,poterya2024uptake} contexts.

In this letter, we investigate with CHGNet the structural properties of two sets of materials involving either
nano-particles or ultra-thin oxides on a surface. Varying system sizes
and stoichiometries, we find for both types of materials that the universal
potential is sufficiently accurate to identify the global minimum
energy structures of some of the systems studied. However, we also
find cases in which CHGNet lacks such accuracy. These cases form
the basis for testing how to improve the pretrained CHGNet (v0.3.0). 
We pursue
two avenues for this. Either we fine-tune the network, or we
add an extra machine learning model to the
potential prediction in what is known as $\Delta$-learning\cite{ramakrishnan2015big}.

The letter is organized in the following manner: firstly we present two
silicate clusters, and augment the CHGNet energy predictions 
where necessary, following
either a fine-tuning or $\Delta$-learning strategy. Next, the effect of
the augmentations is discussed as a function of the amount of data
used. The letter proceeds to consider CHGNet
predictions for ultra-thin oxides on the Ag(111) surface. Augmentation
is found necessary for experimental consistency, 
and the $\Delta$-correction scheme is employed
leading to the construction of a reliable model. Finally, the model is
applied to very large surface defect structures containing more than $1000$
atoms.

Throughout this work, we will utilise optimization methods as
implemented in the "Atomistic Global Optimization X" package,
AGOX(v3.6.0)\cite{AGOX} that builds on the "Atomic Simulation Environment"
(ASE)\cite{larsen2017atomic}. Density Functional Theory (DFT) calculations are performed in GPAW\cite{GPAW}.
More details on the methods used are given in Section S{\sc II} and Section S{\sc IV}\cite{supplementary}.

\begin{figure}[!ht]
  {
    \centering
    \includegraphics[width=1.0\linewidth]{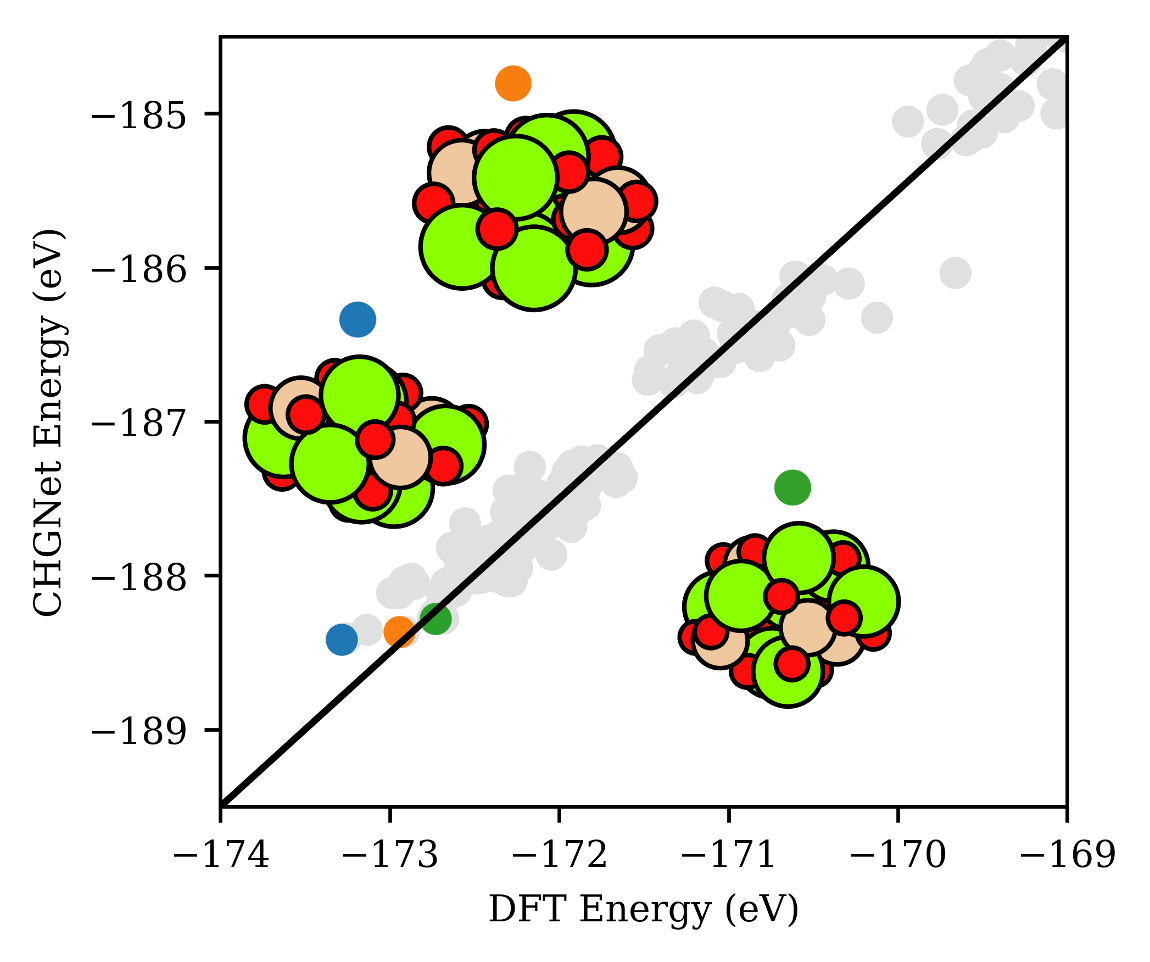}
    }
    \caption{Parity plot of CHGNet prediction vs DFT energy for
      four formula unit Olivine clusters. Three unique low-energy
      configurations are illustrated and corresponding datapoints indicated.
      The solid line is a guide to the eye.
      Green, tan and red spheres represent magnesium, silicon and oxygen atoms, respectively}
    \label{fig:silicates}
\end{figure}

Figure \figref[]{fig:silicates} presents three low energy
configurations found for a silicate nano-particle with four olivine
formula units, i.e.\
$\bigl[\mathrm{Mg}\mathrm{Si}_2\mathrm{O}_4\bigr]_4$. For these configurations,
we find that CHGNet
correctly predicts the same ordering as DFT\cite{slavensky2023generating}.
This is evidenced by the parity plot of CHGNet versus DFT energies presented. We
find similar DFT reproducibility behaviour of CHGNet for a range of other structures, 
see Section S{\sc I}. This result demonstrates that despite
CHGNet being trained largely on extended solid materials, it
contains sufficient chemical insight so as to deal with the chemical
consequences of compounds having finite size whereby many atoms end up
in surface sites.

The offset of CHGNet energy to DFT energies apparent from Fig.\
\figref[]{fig:silicates} stems from the use of different DFT codes and
DFT run-time parameters. The CHGNet architecture comes with 94 parameters,
one for
each atomic species, known as the \textit{composition model}, which
may be altered to accomodate such energy differences. Section S{\sc V}
discusses how to adjust the composition model parameters and
eliminate the offset.

Using CHGNet for a wide variety of materials, we find occasional
examples where it does not provide the same minimum energy
configuration as found with DFT. An example of this is given in
Fig.\ \figref[]{fig:fine-tuning}(a-c) that present three 
low-energy configurations of silicate nano-particles with
four pyroxene formula units, i.e.\
$\bigl[\mathrm{Mg}\mathrm{Si}\mathrm{O}_3\bigr]_4$ (PYR-4). In this case,
the CHGNet model predicts the wrong stability ordering of the
cluster configurations, as seen from the parity plot in Fig.\
\figref[]{fig:fine-tuning}(e). The structure preferred in CHGNet,
Fig.\ \figref[]{fig:fine-tuning}(a), is hollow, while the one 
preferred in DFT\cite{Escatllar2019},
Fig.\ \figref[]{fig:fine-tuning}(c), is more compact. 

To rationalize CHGNet's behavior for this system, we note that it 
relies on extrapolating the total energy of unknown
systems by generalizing from local atomic descriptors in known systems.
These descriptors result from graph-embedding in consecutive
deep neural network layers, and predictions for certain motifs may be
off, if these are not present in
the training data. 

We observe this effect whilst calculating the total dipole moments of these structures, 
which yields 1.22 and $<0.01$ $|e|\cdot\angstrom$, for the CHGNet and DFT minima respectively. 
The much higher value of the CHGNet minimum occurs with the presence of 
magnesium coordinated magnesium in a triangular motif capped by a
single oxygen, cf.\ upper part of structure in Fig.\ \figref[]{fig:fine-tuning}(a).
We theorise that this motif has been incorrectly extrapolated from 
structures in which it correlates with stability.

Postulating thusly, the obvious solution is to augment the model with a
collection of such data. In the following we shall take two approaches
to such augmentation.

We first consider the option of fine-tuning CHGNet \cite{deng2024overcoming,focassio2024performance}. Fine-tuning
involves:
\begin{itemize}
\item
  Acquire $N_\mathrm{sub}$ pieces of data, that specify structures and
  corresponding DFT energies and forces.
\item
  Split the data in training and validation data randomly in some desired
  ratio, e.g.\ 90:10. 
\item
  Complete a training step in which the network parameters are
  updated via backpropagation of the training data.
\item
  Repeat steps until the model prediction accuracy on the validation 
  set cannot be improved any further.
\end{itemize}

\begin{figure*}[ht!]
  {
    \centering
    \includegraphics[width=0.7\textwidth]{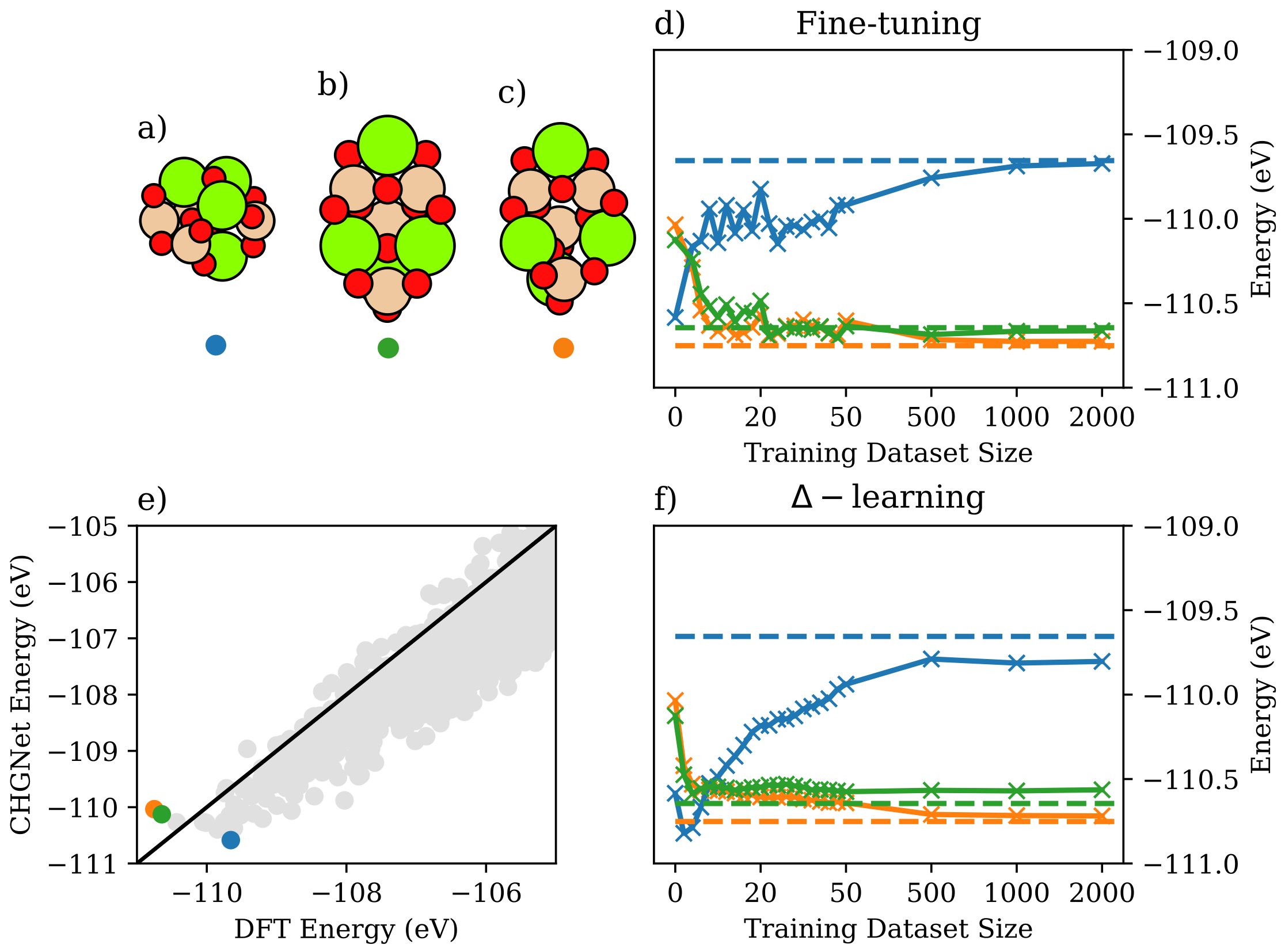}
    }
    \caption{(a-c) depict three low energy cluster configurations of PYR-4. (e) parity plot
    of the CHGNet vs DFT energies. (d) Fine-tuning prediction of energies with increasing data.
    (f) $\Delta-$model prediction of energies with increasing data. Structures (a-c) are color 
    coded in all cases, with dashed lines indicating DFT energies.}  
    \label{fig:fine-tuning}
\end{figure*}

Figure \figref[]{fig:fine-tuning}(d) shows the energy prediction for the
three low-energy cluster configurations of PYR-4 as a function of the amount of
data, $1\le N_\mathrm{sub}\le 2000$, randomly extracted from an established dataset.
Details of this dataset are given in the
Section S{\sc VI}.

As successively more data is used, the fine-tuned CHGNet indeed learns the
relative ordering of the three low-energy cluster configurations, cf Fig.\
\figref[]{fig:fine-tuning}(d). Below 50 datapoints used, the
fine-tuned CHGNet reacts turbulently to new data with predictions
fluctuating in a jagged manner. When these
fluctuations settle for $N_\mathrm{sub}\ge 500$, the fine-tuned CHGNet 
matches well the DFT energies of the respective structures. 

Having seen that it is indeed possibly to fine-tune CHGNet to
learn system specific information and attain DFT-level accuracy,
we now present an alternative approach: $\Delta$-learning. 
The motivation for
developing a new approach stems from the fact that depending on
the dataset size and the number of required epochs,
the augmentation of a neural network via fine-tuning may be highly
computationally costly, demanding expensive GPUs. 

$\Delta$-learning is an established machine learning approach
in which one model learns "as much as it can" and a subsequent model
learns the residual errors. Together, the two models are referred to
as a $\Delta$-model. $\Delta$-learning has previously been successfully applied in the 
domain of molecular\cite{kaser2020reactive} and materials\cite{nandi2021delta} modelling,
amongst others \cite{hu2003combined,balabin2009neural,gillan2013first,ramakrishnan2015big,zaspel2018boosting,chmiela2018towards,sauceda2019molecular,stohr2020accurate}.

In the present context, we opt to use a Gaussian Process Regression
(GPR) model for learning the residuals between the CHGNet
and DFT values. We select GPR as an 
auxiliary model for its performance in the low data regime, both
in terms of speed and reliability. Depending on hyper-parameters, a
GPR may further excel in not correcting the CHGNet predictions
outside the regions of the data used for the GPR training. This is
unlike the situation in which a neural network is used as the auxiliary
model, where corrective predictions could be more far reaching in
impact.

In the $\Delta$-model approach we make predictions according
to:
$$ E_{\Delta\mathrm{-model}}(\mathcal{S}) = E^\mathrm{CHGNet}(\mathcal{S}) + \sum^N_i \Delta E^\mathrm{GPR}(\mathbf{x}_i),$$
for an unknown structure, $\mathcal{S}$, with local atomic
representations $\mathbf{x}_i$.
The GPR model prediction for the
residual can be seen in Section S{\sc III}.

We note, that it is instructive to think of a $\Delta$-model
composed of a fixed NN and a trainable GPR model, as a GPR model with a prior given by
the NN, which is concurrent with the language within our AGOX implementation.

The $\Delta$-model approach does not require the expensive GPU
retraining of the neural network as was the case with the fine-tuning
approach. In practice, we use a local SOAP-descriptor-based sparse GPR
model, see Section S{\sc III} for more details\cite{bartok2013,ronne2022atomistic}. 
It would be equally viable to select any equivalently representative 
descriptor, including the intrinsic CHGNet features.

Figure \figref[]{fig:fine-tuning}(f) presents the prediction of the
GPR-based $\Delta$-model as a function of the amount, $N_\mathrm{sub}$, of
data sub-sampled from the dataset. 
The silicate cluster stabilities
predicted with the GPR-based $\Delta$-model are seen to smoothly 
tend towards the DFT values, i.e.\ they are not 
jagged for $N_\mathrm{sub}\le 50$,
and converge for $N_\mathrm{sub}\ge 500$. The ordering becomes
correct much earlier for the $\Delta$-model, being consistently
so after as little at 10 data points. 
This evidences that the GPR-based auxiliary model is
sufficiently versatile to learn the residuals in the space defined by
the silicate clusters considered. 

At this point, it is instructive to discuss the origin of the 
dataset. The dataset was
constructed in consecutive cycles of structural searches followed by
model augmentation. Further details are given in Section S{\sc VI}. Here it
suffices to highlight that such an active learning acquisition of
training data is only computationally feasible if the model 
augmentation step is fast and stable with small additions of data. 
The differences in consistency between Figure \figref[]{fig:fine-tuning}(d,f)
in the low data regime confirms that
datasets can be more efficiently incorporated with the GPR-based
$\Delta$-model.  This leads to them being overall powerful options when
augmenting the predictive power of CHGNet or similar universal potentials.

We now move to consider an application involving surface reconstructions, where
we will demonstrate the use of the $\Delta$-model approach for system
sizes not computationally accessible for DFT studies. Specifically, we shall consider
three ultra-thin oxide phases that have been observed on Ag(111)\cite{Ag111-O-1,derouin2016thermally}.
The
phases are the $c(4\times 8)$, $c(3 \times 5\sqrt{3})$, and $p(4
\times 5\sqrt{3})$, that have 0.5, 0.4, and 0.375 oxygen/surface Ag
coverages, respectively. For convenience, we shall refer to the three
phases as phase {\sc I}, {\sc II}, and {\sc III}, respectively.

\begin{figure}[ht!]
  {
    \centering
    \includegraphics[]{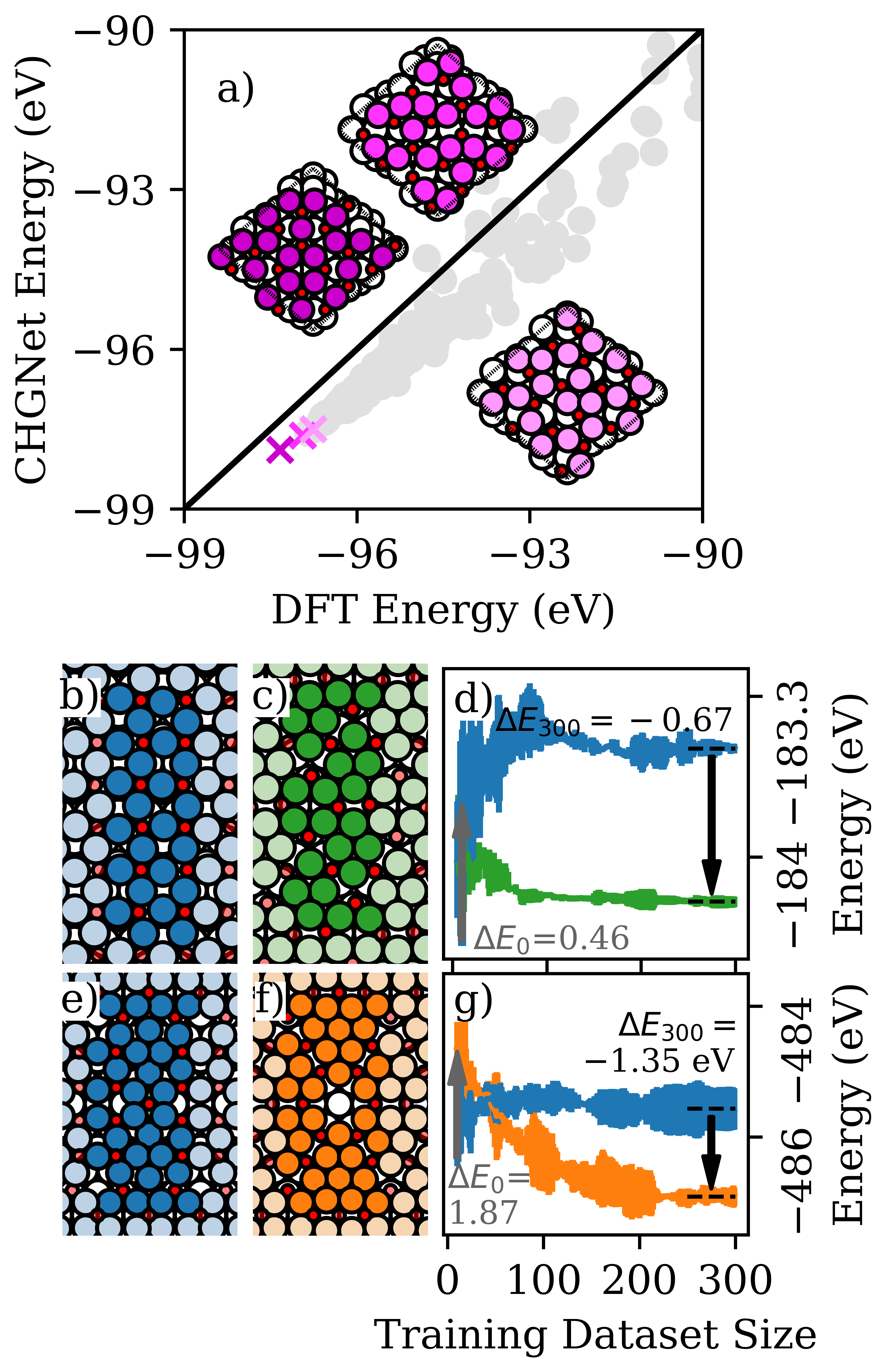}
    }
    \caption{(a) Parity plot of phase {\sc I} structures, with colour coded insets depicting their correspondingly coloured points. DFT and CHGNet agree on the GM. (b) CHGNet GM for phase {\sc
        II}. (c) DFT GM for phase {\sc II}. (d) $\Delta$-model
      predictions for phase {\sc II}. The DFT energy difference is calculated at -0.53 eV.
      (e) CHGNet GM for phase {\sc
        III}. (f) DFT GM for phase {\sc III}. (g) $\Delta$-model
      predictions for phase {\sc III}. The DFT energy difference is -0.76 eV. 
      In all depictions, red spheres represent oxygen, with all others indicating silver.}
    \label{fig:ag111}
\end{figure}
Starting with phase {\sc I}, inserts in Fig. \figref[]{fig:ag111}(a) present
its structure and two alternative structures found when searching in
either the CHGNet or in the full DFT energy landscape. The parity plot
displayed in Fig. \figref[]{fig:ag111}(a) shows that CHGNet works
quite well for this system. Moving to phase {\sc II} in
Fig. \figref[]{fig:ag111}(b,c), the situation changes and now CHGNet
predicts a structure with linear O-Ag-O motifs lining up in rows as
the more stable, Fig.\ \figref[]{fig:ag111}(b), while DFT has an
elaborate structure with Ag$_6$ triangles as the ground states, Fig.\
\figref[]{fig:ag111}(c).

In order to remedy the identified deficiency and to
resolve CHGNets prediction for the correct ground state structure for phase
{\sc II}, we collected, via 10 uncorrelated active learning campaigns,
training data for 10 independent
$\Delta$-models. Figure \figref[]{fig:ag111}(d) shows the mean and uncertainty
of predictions by these 10 models as a function of the amount of training
data used. To put the approach to the test, we collected the data for
the same cell as phase {\sc II} itself, but for different stoichiometries, 
specifically omitting the stoichiometry containing the solution (see SI
for details). Clearly, the models are capable of learning the correct
ordering and energy difference of the two structures.

Finally, in Fig.\ \figref[]{fig:ag111}(e-g) the situation for phase
{\sc III} is presented. Again, the pretrained CHGNet predicts a wrong
ground state structure, Fig.\ \figref[]{fig:ag111}(e), that does not
have the Ag$_6$ and Ag$_{10}$ triangular motifs present in the DFT
ground state structure, Fig.\ \figref[]{fig:ag111}(f). Now
interestingly, the 10 $\Delta$-models that were trained for phase {\sc
  II}, when applied to phase {\sc III} are able to predict the
correct ordering and energy difference, as shown in Fig.\ \figref[]{fig:ag111}(g).
This demonstrates that the
$\Delta$-models generalize well and opens for their application to
more advanced materials problems where DFT calculations become
prohibitively large.

\captionsetup[subfigure]{position=top,textfont=normalfont,singlelinecheck=off}
\begin{figure}[htb!]

    \includegraphics[width=\linewidth]{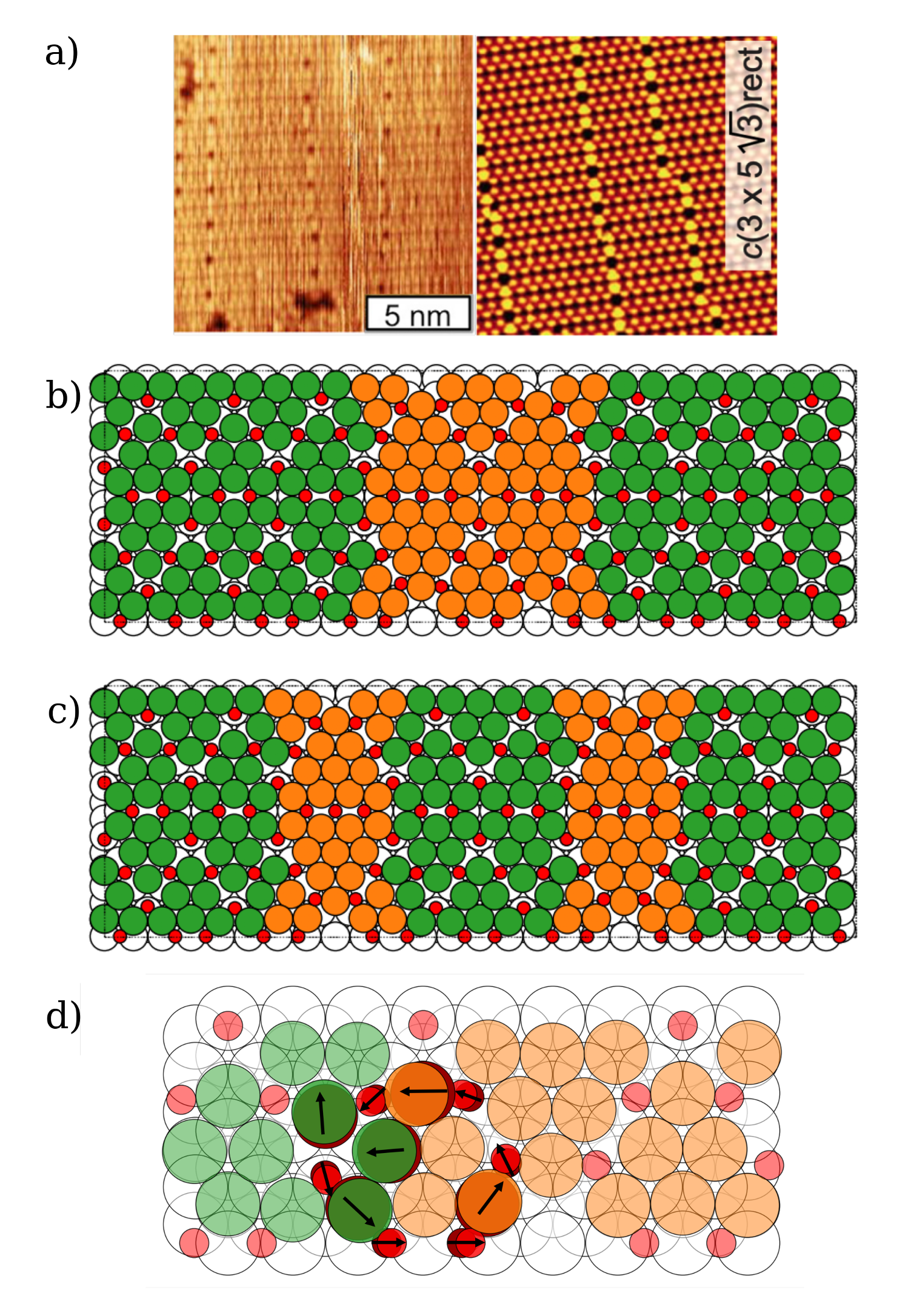}
  
    \caption{a) Reproduction of STM images: Reprinted 
    with permission from Ref.\ \cite{derouin2016thermally}. 
    Copyright 2016 American Chemical Society,  
    and from Ref.\ \cite{Ag111-O-1} under the Creative Commons Attribution 3.0 License,
      respectively. Both images depict the phase {\sc II} ultra-thin
      oxide on Ag(111) under oxygen exposure, where 
      single-width phase {\sc III} stripes can be observed
      consistently. Schematics depicitng b) neighbouring phase {\sc
        III} cells, and c) dispersed {\sc
        III} cells. 
      d) Schematic of the relaxation undergone at the interface between 
      phases {\sc
      II} and {\sc
      III}.}
    \label{fig:large}
\end{figure}

In Fig.\ \figref[]{fig:large} we present such a case. The STM images
in Fig.\ \figref[]{fig:large}(a), obtained independently by the
groups of Refs.\ \onlinecite{Ag111-O-1} and \onlinecite{derouin2016thermally},
suggest the appearance of single rows of
the phase {\sc III} structure inside larger domains of phase {\sc
  II}. These defect rows are observed to move dynamically at room
temperature and their presence therefore cannot be ascribed to kinetic
limitations. To assess the possibility of a thermodynamic origin of the rows we
adopted a $\Delta$-model selected from 
amongst those developed for
Fig.\ \figref[]{fig:ag111}. In Fig.\ \figref[]{fig:large}(b,c) two
systems are presented that explain why for instance the
rows of phase {\sc III} do not collapse to wider domains. In Fig.\
\figref[]{fig:large}(b) two rows of phase {\sc III} rows are adjacent
to each other, and interfaced to phase {\sc II}, while in Fig.\
\figref[]{fig:large}(c) the two rows of phase {\sc III} have dispersed in
to being fully separated by phase {\sc II}. With the augmented CHGNet
model, we can calculate the energy change of this process, and find it
to be preferred by 0.352 eV meaning that the interface energy between
phase {\sc II} and {\sc III} is $-0.176$ eV per $5\sqrt{3}a/\sqrt{2}$ side
length, where $a$ is the lattice constant.

Figure \figref[]{fig:large}(d) shows the considerable
relaxations occurring at the interface between the two phases. 
Omitting the relaxations perpendicular to the interface plane the
interface energy becomes essentially zero, leading us to
conjecture that stress present within phase {\sc II} is relieved at the interface
and constitutes the driving mechanism for the presence of the rows of phase
{\sc III} defects within the large domains of phase {\sc II}. The reason
why the system does not evolve into alternating single rows of
either phase, which would maximize the amount of favorable interfaces,
is that this would also bring the overall oxygen coverage and hence
chemisorption energy gain down, given the difference in O-coverage of
phase II and III.

In conclusion, we have investigated the capability of a universal
potential, CHGNet, in describing selected inorganic clusters and
surfaces. In cases where the accuracy of the plain CHGNet model is
insufficient to correctly order low-energy structures, we demonstrate
that either fine-tuning or $\Delta$-learning offer a means to augment
the model. We finally used a $\Delta$-learning augmented CHGNet model
to identify and rationalize negative energy domain boundaries in the
thin film oxides forming on Ag(111).

We acknowledge support by VILLUM FONDEN through Investigator grant,
project no. 16562, and by the Danish National Research Foundation
through the Center of Excellence “InterCat” (Grant agreement no:
DNRF150).

\bibliography{biblio}

\begin{thebibliography}{84}%
\makeatletter
\providecommand \@ifxundefined [1]{%
 \@ifx{#1\undefined}
}%
\providecommand \@ifnum [1]{%
 \ifnum #1\expandafter \@firstoftwo
 \else \expandafter \@secondoftwo
 \fi
}%
\providecommand \@ifx [1]{%
 \ifx #1\expandafter \@firstoftwo
 \else \expandafter \@secondoftwo
 \fi
}%
\providecommand \natexlab [1]{#1}%
\providecommand \enquote  [1]{``#1''}%
\providecommand \bibnamefont  [1]{#1}%
\providecommand \bibfnamefont [1]{#1}%
\providecommand \citenamefont [1]{#1}%
\providecommand \href@noop [0]{\@secondoftwo}%
\providecommand \href [0]{\begingroup \@sanitize@url \@href}%
\providecommand \@href[1]{\@@startlink{#1}\@@href}%
\providecommand \@@href[1]{\endgroup#1\@@endlink}%
\providecommand \@sanitize@url [0]{\catcode `\\12\catcode `\$12\catcode
  `\&12\catcode `\#12\catcode `\^12\catcode `\_12\catcode `\%12\relax}%
\providecommand \@@startlink[1]{}%
\providecommand \@@endlink[0]{}%
\providecommand \url  [0]{\begingroup\@sanitize@url \@url }%
\providecommand \@url [1]{\endgroup\@href {#1}{\urlprefix }}%
\providecommand \urlprefix  [0]{URL }%
\providecommand \Eprint [0]{\href }%
\providecommand \doibase [0]{http://dx.doi.org/}%
\providecommand \selectlanguage [0]{\@gobble}%
\providecommand \bibinfo  [0]{\@secondoftwo}%
\providecommand \bibfield  [0]{\@secondoftwo}%
\providecommand \translation [1]{[#1]}%
\providecommand \BibitemOpen [0]{}%
\providecommand \bibitemStop [0]{}%
\providecommand \bibitemNoStop [0]{.\EOS\space}%
\providecommand \EOS [0]{\spacefactor3000\relax}%
\providecommand \BibitemShut  [1]{\csname bibitem#1\endcsname}%
\let\auto@bib@innerbib\@empty
\bibitem [{\citenamefont {Deng}\ \emph {et~al.}(2023)\citenamefont {Deng},
  \citenamefont {Zhong}, \citenamefont {Jun}, \citenamefont {Riebesell},
  \citenamefont {Han}, \citenamefont {Bartel},\ and\ \citenamefont
  {Ceder}}]{deng2023chgnet}%
  \BibitemOpen
  \bibfield  {author} {\bibinfo {author} {\bibfnamefont {B.}~\bibnamefont
  {Deng}}, \bibinfo {author} {\bibfnamefont {P.}~\bibnamefont {Zhong}},
  \bibinfo {author} {\bibfnamefont {K.}~\bibnamefont {Jun}}, \bibinfo {author}
  {\bibfnamefont {J.}~\bibnamefont {Riebesell}}, \bibinfo {author}
  {\bibfnamefont {K.}~\bibnamefont {Han}}, \bibinfo {author} {\bibfnamefont
  {C.~J.}\ \bibnamefont {Bartel}}, \ and\ \bibinfo {author} {\bibfnamefont
  {G.}~\bibnamefont {Ceder}},\ }\href@noop {} {\bibfield  {journal} {\bibinfo
  {journal} {Nature Machine Intelligence}\ }\textbf {\bibinfo {volume} {5}},\
  \bibinfo {pages} {1031} (\bibinfo {year} {2023})}\BibitemShut {NoStop}%
\bibitem [{\citenamefont {Fewster}(2004)}]{fewster2004advances}%
  \BibitemOpen
  \bibfield  {author} {\bibinfo {author} {\bibfnamefont {P.~F.}\ \bibnamefont
  {Fewster}},\ }\href@noop {} {\bibfield  {journal} {\bibinfo  {journal}
  {Progress in crystal growth and characterization of materials}\ }\textbf
  {\bibinfo {volume} {48}},\ \bibinfo {pages} {245} (\bibinfo {year}
  {2004})}\BibitemShut {NoStop}%
\bibitem [{\citenamefont {Stobinski}\ \emph {et~al.}(2014)\citenamefont
  {Stobinski}, \citenamefont {Lesiak}, \citenamefont {Malolepszy},
  \citenamefont {Mazurkiewicz}, \citenamefont {Mierzwa}, \citenamefont {Zemek},
  \citenamefont {Jiricek},\ and\ \citenamefont
  {Bieloshapka}}]{stobinski2014graphene}%
  \BibitemOpen
  \bibfield  {author} {\bibinfo {author} {\bibfnamefont {L.}~\bibnamefont
  {Stobinski}}, \bibinfo {author} {\bibfnamefont {B.}~\bibnamefont {Lesiak}},
  \bibinfo {author} {\bibfnamefont {A.}~\bibnamefont {Malolepszy}}, \bibinfo
  {author} {\bibfnamefont {M.}~\bibnamefont {Mazurkiewicz}}, \bibinfo {author}
  {\bibfnamefont {B.}~\bibnamefont {Mierzwa}}, \bibinfo {author} {\bibfnamefont
  {J.}~\bibnamefont {Zemek}}, \bibinfo {author} {\bibfnamefont
  {P.}~\bibnamefont {Jiricek}}, \ and\ \bibinfo {author} {\bibfnamefont
  {I.}~\bibnamefont {Bieloshapka}},\ }\href@noop {} {\bibfield  {journal}
  {\bibinfo  {journal} {Journal of Electron Spectroscopy and Related
  Phenomena}\ }\textbf {\bibinfo {volume} {195}},\ \bibinfo {pages} {145}
  (\bibinfo {year} {2014})}\BibitemShut {NoStop}%
\bibitem [{\citenamefont {{\v{C}}ern{\`y}}(2017)}]{vcerny2017crystal}%
  \BibitemOpen
  \bibfield  {author} {\bibinfo {author} {\bibfnamefont {R.}~\bibnamefont
  {{\v{C}}ern{\`y}}},\ }\href@noop {} {\bibfield  {journal} {\bibinfo
  {journal} {Crystals}\ }\textbf {\bibinfo {volume} {7}},\ \bibinfo {pages}
  {142} (\bibinfo {year} {2017})}\BibitemShut {NoStop}%
\bibitem [{\citenamefont {Gloystein}\ \emph {et~al.}(2020)\citenamefont
  {Gloystein}, \citenamefont {Nilius}, \citenamefont {Goniakowski},\ and\
  \citenamefont {Noguera}}]{gloystein2020nanopyramidal}%
  \BibitemOpen
  \bibfield  {author} {\bibinfo {author} {\bibfnamefont {A.}~\bibnamefont
  {Gloystein}}, \bibinfo {author} {\bibfnamefont {N.}~\bibnamefont {Nilius}},
  \bibinfo {author} {\bibfnamefont {J.}~\bibnamefont {Goniakowski}}, \ and\
  \bibinfo {author} {\bibfnamefont {C.}~\bibnamefont {Noguera}},\ }\href@noop
  {} {\bibfield  {journal} {\bibinfo  {journal} {The Journal of Physical
  Chemistry C}\ }\textbf {\bibinfo {volume} {124}},\ \bibinfo {pages} {26937}
  (\bibinfo {year} {2020})}\BibitemShut {NoStop}%
\bibitem [{\citenamefont {Shields}\ and\ \citenamefont
  {Gupta}(2024)}]{shields2024stm}%
  \BibitemOpen
  \bibfield  {author} {\bibinfo {author} {\bibfnamefont {S.~S.}\ \bibnamefont
  {Shields}}\ and\ \bibinfo {author} {\bibfnamefont {J.~A.}\ \bibnamefont
  {Gupta}},\ }\href@noop {} {\bibfield  {journal} {\bibinfo  {journal} {Surface
  Science}\ }\textbf {\bibinfo {volume} {740}},\ \bibinfo {pages} {122403}
  (\bibinfo {year} {2024})}\BibitemShut {NoStop}%
\bibitem [{\citenamefont {Guti{\'e}rrez}\ \emph {et~al.}(2001)\citenamefont
  {Guti{\'e}rrez}, \citenamefont {Taga},\ and\ \citenamefont
  {Johansson}}]{gutierrez2001theoretical}%
  \BibitemOpen
  \bibfield  {author} {\bibinfo {author} {\bibfnamefont {G.}~\bibnamefont
  {Guti{\'e}rrez}}, \bibinfo {author} {\bibfnamefont {A.}~\bibnamefont {Taga}},
  \ and\ \bibinfo {author} {\bibfnamefont {B.}~\bibnamefont {Johansson}},\
  }\href@noop {} {\bibfield  {journal} {\bibinfo  {journal} {Physical review
  B}\ }\textbf {\bibinfo {volume} {65}},\ \bibinfo {pages} {012101} (\bibinfo
  {year} {2001})}\BibitemShut {NoStop}%
\bibitem [{\citenamefont {Zhang}\ \emph {et~al.}(2018)\citenamefont {Zhang},
  \citenamefont {Li}, \citenamefont {Frazer}, \citenamefont {Chang},
  \citenamefont {Poeppelmeier}, \citenamefont {Chan},\ and\ \citenamefont
  {Guest}}]{zhang2018atomistic}%
  \BibitemOpen
  \bibfield  {author} {\bibinfo {author} {\bibfnamefont {R.}~\bibnamefont
  {Zhang}}, \bibinfo {author} {\bibfnamefont {L.}~\bibnamefont {Li}}, \bibinfo
  {author} {\bibfnamefont {L.}~\bibnamefont {Frazer}}, \bibinfo {author}
  {\bibfnamefont {K.~B.}\ \bibnamefont {Chang}}, \bibinfo {author}
  {\bibfnamefont {K.~R.}\ \bibnamefont {Poeppelmeier}}, \bibinfo {author}
  {\bibfnamefont {M.~K.}\ \bibnamefont {Chan}}, \ and\ \bibinfo {author}
  {\bibfnamefont {J.~R.}\ \bibnamefont {Guest}},\ }\href@noop {} {\bibfield
  {journal} {\bibinfo  {journal} {Physical Chemistry Chemical Physics}\
  }\textbf {\bibinfo {volume} {20}},\ \bibinfo {pages} {27456} (\bibinfo {year}
  {2018})}\BibitemShut {NoStop}%
\bibitem [{\citenamefont {Hendriksen}\ \emph {et~al.}(2010)\citenamefont
  {Hendriksen}, \citenamefont {Ackermann}, \citenamefont {Van~Rijn},
  \citenamefont {Stoltz}, \citenamefont {Popa}, \citenamefont {Balmes},
  \citenamefont {Resta}, \citenamefont {Wermeille}, \citenamefont {Felici},
  \citenamefont {Ferrer} \emph {et~al.}}]{hendriksen2010role}%
  \BibitemOpen
  \bibfield  {author} {\bibinfo {author} {\bibfnamefont {B.~L.}\ \bibnamefont
  {Hendriksen}}, \bibinfo {author} {\bibfnamefont {M.~D.}\ \bibnamefont
  {Ackermann}}, \bibinfo {author} {\bibfnamefont {R.}~\bibnamefont {Van~Rijn}},
  \bibinfo {author} {\bibfnamefont {D.}~\bibnamefont {Stoltz}}, \bibinfo
  {author} {\bibfnamefont {I.}~\bibnamefont {Popa}}, \bibinfo {author}
  {\bibfnamefont {O.}~\bibnamefont {Balmes}}, \bibinfo {author} {\bibfnamefont
  {A.}~\bibnamefont {Resta}}, \bibinfo {author} {\bibfnamefont
  {D.}~\bibnamefont {Wermeille}}, \bibinfo {author} {\bibfnamefont
  {R.}~\bibnamefont {Felici}}, \bibinfo {author} {\bibfnamefont
  {S.}~\bibnamefont {Ferrer}},  \emph {et~al.},\ }\href@noop {} {\bibfield
  {journal} {\bibinfo  {journal} {Nature chemistry}\ }\textbf {\bibinfo
  {volume} {2}},\ \bibinfo {pages} {730} (\bibinfo {year} {2010})}\BibitemShut
  {NoStop}%
\bibitem [{\citenamefont {Polo-Garzon}\ \emph {et~al.}(2019)\citenamefont
  {Polo-Garzon}, \citenamefont {Bao}, \citenamefont {Zhang}, \citenamefont
  {Huang},\ and\ \citenamefont {Wu}}]{polo2019surface}%
  \BibitemOpen
  \bibfield  {author} {\bibinfo {author} {\bibfnamefont {F.}~\bibnamefont
  {Polo-Garzon}}, \bibinfo {author} {\bibfnamefont {Z.}~\bibnamefont {Bao}},
  \bibinfo {author} {\bibfnamefont {X.}~\bibnamefont {Zhang}}, \bibinfo
  {author} {\bibfnamefont {W.}~\bibnamefont {Huang}}, \ and\ \bibinfo {author}
  {\bibfnamefont {Z.}~\bibnamefont {Wu}},\ }\href@noop {} {\bibfield  {journal}
  {\bibinfo  {journal} {Acs Catalysis}\ }\textbf {\bibinfo {volume} {9}},\
  \bibinfo {pages} {5692} (\bibinfo {year} {2019})}\BibitemShut {NoStop}%
\bibitem [{\citenamefont {Chen}\ \emph {et~al.}(2019)\citenamefont {Chen},
  \citenamefont {Xiong},\ and\ \citenamefont {Huang}}]{chen2019surface}%
  \BibitemOpen
  \bibfield  {author} {\bibinfo {author} {\bibfnamefont {S.}~\bibnamefont
  {Chen}}, \bibinfo {author} {\bibfnamefont {F.}~\bibnamefont {Xiong}}, \ and\
  \bibinfo {author} {\bibfnamefont {W.}~\bibnamefont {Huang}},\ }\href@noop {}
  {\bibfield  {journal} {\bibinfo  {journal} {Surface Science Reports}\
  }\textbf {\bibinfo {volume} {74}},\ \bibinfo {pages} {100471} (\bibinfo
  {year} {2019})}\BibitemShut {NoStop}%
\bibitem [{\citenamefont {Baker}\ \emph {et~al.}(2022)\citenamefont {Baker},
  \citenamefont {Pitfield}, \citenamefont {Price},\ and\ \citenamefont
  {Hepplestone}}]{Baker2022}%
  \BibitemOpen
  \bibfield  {author} {\bibinfo {author} {\bibfnamefont {E.~A.~D.}\
  \bibnamefont {Baker}}, \bibinfo {author} {\bibfnamefont {J.}~\bibnamefont
  {Pitfield}}, \bibinfo {author} {\bibfnamefont {C.~J.}\ \bibnamefont {Price}},
  \ and\ \bibinfo {author} {\bibfnamefont {S.~P.}\ \bibnamefont
  {Hepplestone}},\ }\href {\doibase 10.1088/1361-648X/ac7d2c} {\bibfield
  {journal} {\bibinfo  {journal} {Journal of Physics: Condensed Matter}\
  }\textbf {\bibinfo {volume} {34}},\ \bibinfo {pages} {375001} (\bibinfo
  {year} {2022})}\BibitemShut {NoStop}%
\bibitem [{\citenamefont {Hill}\ \emph {et~al.}(2000)\citenamefont {Hill},
  \citenamefont {Beach},\ and\ \citenamefont {McGill}}]{hill2000nickel}%
  \BibitemOpen
  \bibfield  {author} {\bibinfo {author} {\bibfnamefont {C.}~\bibnamefont
  {Hill}}, \bibinfo {author} {\bibfnamefont {R.}~\bibnamefont {Beach}}, \ and\
  \bibinfo {author} {\bibfnamefont {T.}~\bibnamefont {McGill}},\ }\href
  {\doibase https://doi.org/10.1116/1.1306283} {\bibfield  {journal} {\bibinfo
  {journal} {Journal of Vacuum Science \& Technology B: Microelectronics and
  Nanometer Structures Processing, Measurement, and Phenomena}\ }\textbf
  {\bibinfo {volume} {18}},\ \bibinfo {pages} {2044} (\bibinfo {year}
  {2000})}\BibitemShut {NoStop}%
\bibitem [{\citenamefont {Schusteritsch}\ \emph {et~al.}(2015)\citenamefont
  {Schusteritsch}, \citenamefont {Hepplestone},\ and\ \citenamefont
  {Pickard}}]{schusteritsch2015first}%
  \BibitemOpen
  \bibfield  {author} {\bibinfo {author} {\bibfnamefont {G.}~\bibnamefont
  {Schusteritsch}}, \bibinfo {author} {\bibfnamefont {S.~P.}\ \bibnamefont
  {Hepplestone}}, \ and\ \bibinfo {author} {\bibfnamefont {C.~J.}\ \bibnamefont
  {Pickard}},\ }\href@noop {} {\bibfield  {journal} {\bibinfo  {journal}
  {Physical Review B}\ }\textbf {\bibinfo {volume} {92}},\ \bibinfo {pages}
  {054105} (\bibinfo {year} {2015})}\BibitemShut {NoStop}%
\bibitem [{\citenamefont {Taylor}\ \emph {et~al.}(2020)\citenamefont {Taylor},
  \citenamefont {Davies}, \citenamefont {Rudkin}, \citenamefont {Price},
  \citenamefont {Chan},\ and\ \citenamefont {Hepplestone}}]{TAYLOR2020107515}%
  \BibitemOpen
  \bibfield  {author} {\bibinfo {author} {\bibfnamefont {N.~T.}\ \bibnamefont
  {Taylor}}, \bibinfo {author} {\bibfnamefont {F.~H.}\ \bibnamefont {Davies}},
  \bibinfo {author} {\bibfnamefont {I.~E.~M.}\ \bibnamefont {Rudkin}}, \bibinfo
  {author} {\bibfnamefont {C.~J.}\ \bibnamefont {Price}}, \bibinfo {author}
  {\bibfnamefont {T.~H.}\ \bibnamefont {Chan}}, \ and\ \bibinfo {author}
  {\bibfnamefont {S.~P.}\ \bibnamefont {Hepplestone}},\ }\href {\doibase
  https://doi.org/10.1016/j.cpc.2020.107515} {\bibfield  {journal} {\bibinfo
  {journal} {Computer Physics Communications}\ }\textbf {\bibinfo {volume}
  {257}},\ \bibinfo {pages} {107515} (\bibinfo {year} {2020})}\BibitemShut
  {NoStop}%
\bibitem [{\citenamefont {Pitfield}\ \emph {et~al.}(2024)\citenamefont
  {Pitfield}, \citenamefont {Taylor},\ and\ \citenamefont
  {Hepplestone}}]{pitfield2024predicting}%
  \BibitemOpen
  \bibfield  {author} {\bibinfo {author} {\bibfnamefont {J.}~\bibnamefont
  {Pitfield}}, \bibinfo {author} {\bibfnamefont {N.}~\bibnamefont {Taylor}}, \
  and\ \bibinfo {author} {\bibfnamefont {S.}~\bibnamefont {Hepplestone}},\
  }\href@noop {} {\bibfield  {journal} {\bibinfo  {journal} {Physical Review
  Letters}\ }\textbf {\bibinfo {volume} {132}},\ \bibinfo {pages} {066201}
  (\bibinfo {year} {2024})}\BibitemShut {NoStop}%
\bibitem [{\citenamefont {Mohr}\ \emph {et~al.}(2015)\citenamefont {Mohr},
  \citenamefont {Ratcliff}, \citenamefont {Genovese}, \citenamefont {Caliste},
  \citenamefont {Boulanger}, \citenamefont {Goedecker},\ and\ \citenamefont
  {Deutsch}}]{mohr2015accurate}%
  \BibitemOpen
  \bibfield  {author} {\bibinfo {author} {\bibfnamefont {S.}~\bibnamefont
  {Mohr}}, \bibinfo {author} {\bibfnamefont {L.~E.}\ \bibnamefont {Ratcliff}},
  \bibinfo {author} {\bibfnamefont {L.}~\bibnamefont {Genovese}}, \bibinfo
  {author} {\bibfnamefont {D.}~\bibnamefont {Caliste}}, \bibinfo {author}
  {\bibfnamefont {P.}~\bibnamefont {Boulanger}}, \bibinfo {author}
  {\bibfnamefont {S.}~\bibnamefont {Goedecker}}, \ and\ \bibinfo {author}
  {\bibfnamefont {T.}~\bibnamefont {Deutsch}},\ }\href@noop {} {\bibfield
  {journal} {\bibinfo  {journal} {Physical Chemistry Chemical Physics}\
  }\textbf {\bibinfo {volume} {17}},\ \bibinfo {pages} {31360} (\bibinfo {year}
  {2015})}\BibitemShut {NoStop}%
\bibitem [{\citenamefont {Behler}\ and\ \citenamefont
  {Parrinello}(2007)}]{behler2007generalized}%
  \BibitemOpen
  \bibfield  {author} {\bibinfo {author} {\bibfnamefont {J.}~\bibnamefont
  {Behler}}\ and\ \bibinfo {author} {\bibfnamefont {M.}~\bibnamefont
  {Parrinello}},\ }\href@noop {} {\bibfield  {journal} {\bibinfo  {journal}
  {Physical review letters}\ }\textbf {\bibinfo {volume} {98}},\ \bibinfo
  {pages} {146401} (\bibinfo {year} {2007})}\BibitemShut {NoStop}%
\bibitem [{\citenamefont {Sch{\"u}tt}\ \emph {et~al.}(2018)\citenamefont
  {Sch{\"u}tt}, \citenamefont {Sauceda}, \citenamefont {Kindermans},
  \citenamefont {Tkatchenko},\ and\ \citenamefont
  {M{\"u}ller}}]{schutt2018schnet}%
  \BibitemOpen
  \bibfield  {author} {\bibinfo {author} {\bibfnamefont {K.~T.}\ \bibnamefont
  {Sch{\"u}tt}}, \bibinfo {author} {\bibfnamefont {H.~E.}\ \bibnamefont
  {Sauceda}}, \bibinfo {author} {\bibfnamefont {P.-J.}\ \bibnamefont
  {Kindermans}}, \bibinfo {author} {\bibfnamefont {A.}~\bibnamefont
  {Tkatchenko}}, \ and\ \bibinfo {author} {\bibfnamefont {K.-R.}\ \bibnamefont
  {M{\"u}ller}},\ }\href@noop {} {\bibfield  {journal} {\bibinfo  {journal}
  {The Journal of Chemical Physics}\ }\textbf {\bibinfo {volume} {148}}
  (\bibinfo {year} {2018})}\BibitemShut {NoStop}%
\bibitem [{\citenamefont {Unke}\ and\ \citenamefont
  {Meuwly}(2019)}]{unke2019physnet}%
  \BibitemOpen
  \bibfield  {author} {\bibinfo {author} {\bibfnamefont {O.~T.}\ \bibnamefont
  {Unke}}\ and\ \bibinfo {author} {\bibfnamefont {M.}~\bibnamefont {Meuwly}},\
  }\href@noop {} {\bibfield  {journal} {\bibinfo  {journal} {Journal of
  chemical theory and computation}\ }\textbf {\bibinfo {volume} {15}},\
  \bibinfo {pages} {3678} (\bibinfo {year} {2019})}\BibitemShut {NoStop}%
\bibitem [{\citenamefont {Gasteiger}\ \emph {et~al.}(2021)\citenamefont
  {Gasteiger}, \citenamefont {Becker},\ and\ \citenamefont
  {G{\"u}nnemann}}]{gasteiger2021gemnet}%
  \BibitemOpen
  \bibfield  {author} {\bibinfo {author} {\bibfnamefont {J.}~\bibnamefont
  {Gasteiger}}, \bibinfo {author} {\bibfnamefont {F.}~\bibnamefont {Becker}}, \
  and\ \bibinfo {author} {\bibfnamefont {S.}~\bibnamefont {G{\"u}nnemann}},\
  }\href@noop {} {\bibfield  {journal} {\bibinfo  {journal} {Advances in Neural
  Information Processing Systems}\ }\textbf {\bibinfo {volume} {34}},\ \bibinfo
  {pages} {6790} (\bibinfo {year} {2021})}\BibitemShut {NoStop}%
\bibitem [{\citenamefont {Sch{\"u}tt}\ \emph {et~al.}(2021)\citenamefont
  {Sch{\"u}tt}, \citenamefont {Unke},\ and\ \citenamefont
  {Gastegger}}]{schutt2021equivariant}%
  \BibitemOpen
  \bibfield  {author} {\bibinfo {author} {\bibfnamefont {K.}~\bibnamefont
  {Sch{\"u}tt}}, \bibinfo {author} {\bibfnamefont {O.}~\bibnamefont {Unke}}, \
  and\ \bibinfo {author} {\bibfnamefont {M.}~\bibnamefont {Gastegger}},\ }in\
  \href@noop {} {\emph {\bibinfo {booktitle} {International Conference on
  Machine Learning}}}\ (\bibinfo {organization} {PMLR},\ \bibinfo {year}
  {2021})\ pp.\ \bibinfo {pages} {9377--9388}\BibitemShut {NoStop}%
\bibitem [{\citenamefont {Batatia}\ \emph {et~al.}(2022)\citenamefont
  {Batatia}, \citenamefont {Kovacs}, \citenamefont {Simm}, \citenamefont
  {Ortner},\ and\ \citenamefont {Cs{\'a}nyi}}]{batatia2022mace}%
  \BibitemOpen
  \bibfield  {author} {\bibinfo {author} {\bibfnamefont {I.}~\bibnamefont
  {Batatia}}, \bibinfo {author} {\bibfnamefont {D.~P.}\ \bibnamefont {Kovacs}},
  \bibinfo {author} {\bibfnamefont {G.}~\bibnamefont {Simm}}, \bibinfo {author}
  {\bibfnamefont {C.}~\bibnamefont {Ortner}}, \ and\ \bibinfo {author}
  {\bibfnamefont {G.}~\bibnamefont {Cs{\'a}nyi}},\ }\href@noop {} {\bibfield
  {journal} {\bibinfo  {journal} {Advances in Neural Information Processing
  Systems}\ }\textbf {\bibinfo {volume} {35}},\ \bibinfo {pages} {11423}
  (\bibinfo {year} {2022})}\BibitemShut {NoStop}%
\bibitem [{\citenamefont {Batzner}\ \emph {et~al.}(2022)\citenamefont
  {Batzner}, \citenamefont {Musaelian}, \citenamefont {Sun}, \citenamefont
  {Geiger}, \citenamefont {Mailoa}, \citenamefont {Kornbluth}, \citenamefont
  {Molinari}, \citenamefont {Smidt},\ and\ \citenamefont
  {Kozinsky}}]{batzner20223}%
  \BibitemOpen
  \bibfield  {author} {\bibinfo {author} {\bibfnamefont {S.}~\bibnamefont
  {Batzner}}, \bibinfo {author} {\bibfnamefont {A.}~\bibnamefont {Musaelian}},
  \bibinfo {author} {\bibfnamefont {L.}~\bibnamefont {Sun}}, \bibinfo {author}
  {\bibfnamefont {M.}~\bibnamefont {Geiger}}, \bibinfo {author} {\bibfnamefont
  {J.~P.}\ \bibnamefont {Mailoa}}, \bibinfo {author} {\bibfnamefont
  {M.}~\bibnamefont {Kornbluth}}, \bibinfo {author} {\bibfnamefont
  {N.}~\bibnamefont {Molinari}}, \bibinfo {author} {\bibfnamefont {T.~E.}\
  \bibnamefont {Smidt}}, \ and\ \bibinfo {author} {\bibfnamefont
  {B.}~\bibnamefont {Kozinsky}},\ }\href {\doibase 10.1038/s41467-022-29939-5}
  {\bibfield  {journal} {\bibinfo  {journal} {Nature communications}\ }\textbf
  {\bibinfo {volume} {13}},\ \bibinfo {pages} {2453} (\bibinfo {year}
  {2022})}\BibitemShut {NoStop}%
\bibitem [{\citenamefont {Denzel}\ and\ \citenamefont
  {K{\"a}stner}(2018)}]{denzel2018gaussian}%
  \BibitemOpen
  \bibfield  {author} {\bibinfo {author} {\bibfnamefont {A.}~\bibnamefont
  {Denzel}}\ and\ \bibinfo {author} {\bibfnamefont {J.}~\bibnamefont
  {K{\"a}stner}},\ }\href@noop {} {\bibfield  {journal} {\bibinfo  {journal}
  {The Journal of Chemical Physics}\ }\textbf {\bibinfo {volume} {148}}
  (\bibinfo {year} {2018})}\BibitemShut {NoStop}%
\bibitem [{\citenamefont {Koistinen}\ \emph
  {et~al.}(2019{\natexlab{a}})\citenamefont {Koistinen}, \citenamefont
  {{\'A}sgeirsson}, \citenamefont {Vehtari},\ and\ \citenamefont
  {J{\'o}nsson}}]{koistinen2019minimum}%
  \BibitemOpen
  \bibfield  {author} {\bibinfo {author} {\bibfnamefont {O.-P.}\ \bibnamefont
  {Koistinen}}, \bibinfo {author} {\bibfnamefont {V.}~\bibnamefont
  {{\'A}sgeirsson}}, \bibinfo {author} {\bibfnamefont {A.}~\bibnamefont
  {Vehtari}}, \ and\ \bibinfo {author} {\bibfnamefont {H.}~\bibnamefont
  {J{\'o}nsson}},\ }\href {\doibase https://doi.org/10.1021/acs.jctc.9b01038}
  {\bibfield  {journal} {\bibinfo  {journal} {Journal of Chemical Theory and
  Computation}\ }\textbf {\bibinfo {volume} {16}},\ \bibinfo {pages} {499}
  (\bibinfo {year} {2019}{\natexlab{a}})}\BibitemShut {NoStop}%
\bibitem [{\citenamefont {Koistinen}\ \emph
  {et~al.}(2019{\natexlab{b}})\citenamefont {Koistinen}, \citenamefont
  {{\'A}sgeirsson}, \citenamefont {Vehtari},\ and\ \citenamefont
  {J{\'o}nsson}}]{koistinen2019nudged}%
  \BibitemOpen
  \bibfield  {author} {\bibinfo {author} {\bibfnamefont {O.-P.}\ \bibnamefont
  {Koistinen}}, \bibinfo {author} {\bibfnamefont {V.}~\bibnamefont
  {{\'A}sgeirsson}}, \bibinfo {author} {\bibfnamefont {A.}~\bibnamefont
  {Vehtari}}, \ and\ \bibinfo {author} {\bibfnamefont {H.}~\bibnamefont
  {J{\'o}nsson}},\ }\href {\doibase https://doi.org/10.1021/acs.jctc.9b00692}
  {\bibfield  {journal} {\bibinfo  {journal} {Journal of chemical theory and
  computation}\ }\textbf {\bibinfo {volume} {15}},\ \bibinfo {pages} {6738}
  (\bibinfo {year} {2019}{\natexlab{b}})}\BibitemShut {NoStop}%
\bibitem [{\citenamefont {Deringer}\ \emph {et~al.}(2021)\citenamefont
  {Deringer}, \citenamefont {Bart{\'o}k}, \citenamefont {Bernstein},
  \citenamefont {Wilkins}, \citenamefont {Ceriotti},\ and\ \citenamefont
  {Cs{\'a}nyi}}]{deringer2021gaussian}%
  \BibitemOpen
  \bibfield  {author} {\bibinfo {author} {\bibfnamefont {V.~L.}\ \bibnamefont
  {Deringer}}, \bibinfo {author} {\bibfnamefont {A.~P.}\ \bibnamefont
  {Bart{\'o}k}}, \bibinfo {author} {\bibfnamefont {N.}~\bibnamefont
  {Bernstein}}, \bibinfo {author} {\bibfnamefont {D.~M.}\ \bibnamefont
  {Wilkins}}, \bibinfo {author} {\bibfnamefont {M.}~\bibnamefont {Ceriotti}}, \
  and\ \bibinfo {author} {\bibfnamefont {G.}~\bibnamefont {Cs{\'a}nyi}},\
  }\href@noop {} {\bibfield  {journal} {\bibinfo  {journal} {Chemical Reviews}\
  }\textbf {\bibinfo {volume} {121}},\ \bibinfo {pages} {10073} (\bibinfo
  {year} {2021})}\BibitemShut {NoStop}%
\bibitem [{\citenamefont {Bisbo}\ and\ \citenamefont
  {Hammer}(2022)}]{bisbo2022global}%
  \BibitemOpen
  \bibfield  {author} {\bibinfo {author} {\bibfnamefont {M.~K.}\ \bibnamefont
  {Bisbo}}\ and\ \bibinfo {author} {\bibfnamefont {B.}~\bibnamefont {Hammer}},\
  }\href {\doibase 10.1103/PhysRevB.105.245404} {\bibfield  {journal} {\bibinfo
   {journal} {Physical Review B}\ }\textbf {\bibinfo {volume} {105}},\ \bibinfo
  {pages} {245404} (\bibinfo {year} {2022})}\BibitemShut {NoStop}%
\bibitem [{\citenamefont {Merte}\ \emph {et~al.}(2022)\citenamefont {Merte},
  \citenamefont {Bisbo}, \citenamefont {Sokolovi{\'c}}, \citenamefont
  {Setv{\'\i}n}, \citenamefont {Hagman}, \citenamefont {Shipilin},
  \citenamefont {Schmid}, \citenamefont {Diebold}, \citenamefont {Lundgren},\
  and\ \citenamefont {Hammer}}]{merte2022structure}%
  \BibitemOpen
  \bibfield  {author} {\bibinfo {author} {\bibfnamefont {L.~R.}\ \bibnamefont
  {Merte}}, \bibinfo {author} {\bibfnamefont {M.~K.}\ \bibnamefont {Bisbo}},
  \bibinfo {author} {\bibfnamefont {I.}~\bibnamefont {Sokolovi{\'c}}}, \bibinfo
  {author} {\bibfnamefont {M.}~\bibnamefont {Setv{\'\i}n}}, \bibinfo {author}
  {\bibfnamefont {B.}~\bibnamefont {Hagman}}, \bibinfo {author} {\bibfnamefont
  {M.}~\bibnamefont {Shipilin}}, \bibinfo {author} {\bibfnamefont
  {M.}~\bibnamefont {Schmid}}, \bibinfo {author} {\bibfnamefont
  {U.}~\bibnamefont {Diebold}}, \bibinfo {author} {\bibfnamefont
  {E.}~\bibnamefont {Lundgren}}, \ and\ \bibinfo {author} {\bibfnamefont
  {B.}~\bibnamefont {Hammer}},\ }\href@noop {} {\bibfield  {journal} {\bibinfo
  {journal} {Angewandte Chemie}\ }\textbf {\bibinfo {volume} {134}},\ \bibinfo
  {pages} {e202204244} (\bibinfo {year} {2022})}\BibitemShut {NoStop}%
\bibitem [{\citenamefont {Hamamoto}\ \emph {et~al.}(2023)\citenamefont
  {Hamamoto}, \citenamefont {Pham}, \citenamefont {Bisbo}, \citenamefont
  {Hammer},\ and\ \citenamefont {Morikawa}}]{hamamoto2023machine}%
  \BibitemOpen
  \bibfield  {author} {\bibinfo {author} {\bibfnamefont {Y.}~\bibnamefont
  {Hamamoto}}, \bibinfo {author} {\bibfnamefont {T.~N.}\ \bibnamefont {Pham}},
  \bibinfo {author} {\bibfnamefont {M.~K.}\ \bibnamefont {Bisbo}}, \bibinfo
  {author} {\bibfnamefont {B.}~\bibnamefont {Hammer}}, \ and\ \bibinfo {author}
  {\bibfnamefont {Y.}~\bibnamefont {Morikawa}},\ }\href@noop {} {\bibfield
  {journal} {\bibinfo  {journal} {Physical Review Materials}\ }\textbf
  {\bibinfo {volume} {7}},\ \bibinfo {pages} {124002} (\bibinfo {year}
  {2023})}\BibitemShut {NoStop}%
\bibitem [{\citenamefont {J{\o}rgensen}\ \emph {et~al.}(2019)\citenamefont
  {J{\o}rgensen}, \citenamefont {Mortensen}, \citenamefont {Meldgaard},
  \citenamefont {Kolsbjerg}, \citenamefont {Jacobsen}, \citenamefont
  {S{\o}rensen},\ and\ \citenamefont {Hammer}}]{jorgensen2019atomistic}%
  \BibitemOpen
  \bibfield  {author} {\bibinfo {author} {\bibfnamefont {M.~S.}\ \bibnamefont
  {J{\o}rgensen}}, \bibinfo {author} {\bibfnamefont {H.~L.}\ \bibnamefont
  {Mortensen}}, \bibinfo {author} {\bibfnamefont {S.~A.}\ \bibnamefont
  {Meldgaard}}, \bibinfo {author} {\bibfnamefont {E.~L.}\ \bibnamefont
  {Kolsbjerg}}, \bibinfo {author} {\bibfnamefont {T.~L.}\ \bibnamefont
  {Jacobsen}}, \bibinfo {author} {\bibfnamefont {K.~H.}\ \bibnamefont
  {S{\o}rensen}}, \ and\ \bibinfo {author} {\bibfnamefont {B.}~\bibnamefont
  {Hammer}},\ }\href@noop {} {\bibfield  {journal} {\bibinfo  {journal} {The
  Journal of Chemical Physics}\ }\textbf {\bibinfo {volume} {151}} (\bibinfo
  {year} {2019})}\BibitemShut {NoStop}%
\bibitem [{\citenamefont {Zhou}\ \emph {et~al.}(2019)\citenamefont {Zhou},
  \citenamefont {Kearnes}, \citenamefont {Li}, \citenamefont {Zare},\ and\
  \citenamefont {Riley}}]{zhou2019optimization}%
  \BibitemOpen
  \bibfield  {author} {\bibinfo {author} {\bibfnamefont {Z.}~\bibnamefont
  {Zhou}}, \bibinfo {author} {\bibfnamefont {S.}~\bibnamefont {Kearnes}},
  \bibinfo {author} {\bibfnamefont {L.}~\bibnamefont {Li}}, \bibinfo {author}
  {\bibfnamefont {R.~N.}\ \bibnamefont {Zare}}, \ and\ \bibinfo {author}
  {\bibfnamefont {P.}~\bibnamefont {Riley}},\ }\href {\doibase
  10.1038/s41598-019-47148-x} {\bibfield  {journal} {\bibinfo  {journal}
  {Scientific reports}\ }\textbf {\bibinfo {volume} {9}},\ \bibinfo {pages}
  {10752} (\bibinfo {year} {2019})}\BibitemShut {NoStop}%
\bibitem [{\citenamefont {Gogineni}\ \emph {et~al.}(2020)\citenamefont
  {Gogineni}, \citenamefont {Xu}, \citenamefont {Punzalan}, \citenamefont
  {Jiang}, \citenamefont {Kammeraad}, \citenamefont {Tewari},\ and\
  \citenamefont {Zimmerman}}]{gogineni2020torsionnet}%
  \BibitemOpen
  \bibfield  {author} {\bibinfo {author} {\bibfnamefont {T.}~\bibnamefont
  {Gogineni}}, \bibinfo {author} {\bibfnamefont {Z.}~\bibnamefont {Xu}},
  \bibinfo {author} {\bibfnamefont {E.}~\bibnamefont {Punzalan}}, \bibinfo
  {author} {\bibfnamefont {R.}~\bibnamefont {Jiang}}, \bibinfo {author}
  {\bibfnamefont {J.}~\bibnamefont {Kammeraad}}, \bibinfo {author}
  {\bibfnamefont {A.}~\bibnamefont {Tewari}}, \ and\ \bibinfo {author}
  {\bibfnamefont {P.}~\bibnamefont {Zimmerman}},\ }\href@noop {} {\bibfield
  {journal} {\bibinfo  {journal} {Advances in Neural Information Processing
  Systems}\ }\textbf {\bibinfo {volume} {33}},\ \bibinfo {pages} {20142}
  (\bibinfo {year} {2020})}\BibitemShut {NoStop}%
\bibitem [{\citenamefont {Smith}\ \emph {et~al.}(2018)\citenamefont {Smith},
  \citenamefont {Nebgen}, \citenamefont {Lubbers}, \citenamefont {Isayev},\
  and\ \citenamefont {Roitberg}}]{smith2018less}%
  \BibitemOpen
  \bibfield  {author} {\bibinfo {author} {\bibfnamefont {J.~S.}\ \bibnamefont
  {Smith}}, \bibinfo {author} {\bibfnamefont {B.}~\bibnamefont {Nebgen}},
  \bibinfo {author} {\bibfnamefont {N.}~\bibnamefont {Lubbers}}, \bibinfo
  {author} {\bibfnamefont {O.}~\bibnamefont {Isayev}}, \ and\ \bibinfo {author}
  {\bibfnamefont {A.~E.}\ \bibnamefont {Roitberg}},\ }\href {\doibase
  https://doi.org/10.1063/1.5023802} {\bibfield  {journal} {\bibinfo  {journal}
  {The Journal of chemical physics}\ }\textbf {\bibinfo {volume} {148}},\
  \bibinfo {pages} {241733} (\bibinfo {year} {2018})}\BibitemShut {NoStop}%
\bibitem [{\citenamefont {Persson}\ \emph {et~al.}(2012)\citenamefont
  {Persson}, \citenamefont {Waldwick}, \citenamefont {Lazic},\ and\
  \citenamefont {Ceder}}]{persson2012prediction}%
  \BibitemOpen
  \bibfield  {author} {\bibinfo {author} {\bibfnamefont {K.~A.}\ \bibnamefont
  {Persson}}, \bibinfo {author} {\bibfnamefont {B.}~\bibnamefont {Waldwick}},
  \bibinfo {author} {\bibfnamefont {P.}~\bibnamefont {Lazic}}, \ and\ \bibinfo
  {author} {\bibfnamefont {G.}~\bibnamefont {Ceder}},\ }\href@noop {}
  {\bibfield  {journal} {\bibinfo  {journal} {Physical Review B}\ }\textbf
  {\bibinfo {volume} {85}},\ \bibinfo {pages} {235438} (\bibinfo {year}
  {2012})}\BibitemShut {NoStop}%
\bibitem [{\citenamefont {Chanussot}\ \emph {et~al.}(2021)\citenamefont
  {Chanussot}, \citenamefont {Das}, \citenamefont {Goyal}, \citenamefont
  {Lavril}, \citenamefont {Shuaibi}, \citenamefont {Riviere}, \citenamefont
  {Tran}, \citenamefont {Heras-Domingo}, \citenamefont {Ho}, \citenamefont {Hu}
  \emph {et~al.}}]{chanussot2021open}%
  \BibitemOpen
  \bibfield  {author} {\bibinfo {author} {\bibfnamefont {L.}~\bibnamefont
  {Chanussot}}, \bibinfo {author} {\bibfnamefont {A.}~\bibnamefont {Das}},
  \bibinfo {author} {\bibfnamefont {S.}~\bibnamefont {Goyal}}, \bibinfo
  {author} {\bibfnamefont {T.}~\bibnamefont {Lavril}}, \bibinfo {author}
  {\bibfnamefont {M.}~\bibnamefont {Shuaibi}}, \bibinfo {author} {\bibfnamefont
  {M.}~\bibnamefont {Riviere}}, \bibinfo {author} {\bibfnamefont
  {K.}~\bibnamefont {Tran}}, \bibinfo {author} {\bibfnamefont {J.}~\bibnamefont
  {Heras-Domingo}}, \bibinfo {author} {\bibfnamefont {C.}~\bibnamefont {Ho}},
  \bibinfo {author} {\bibfnamefont {W.}~\bibnamefont {Hu}},  \emph {et~al.},\
  }\href@noop {} {\bibfield  {journal} {\bibinfo  {journal} {Acs Catalysis}\
  }\textbf {\bibinfo {volume} {11}},\ \bibinfo {pages} {6059} (\bibinfo {year}
  {2021})}\BibitemShut {NoStop}%
\bibitem [{\citenamefont {Kim}\ \emph {et~al.}(2016)\citenamefont {Kim},
  \citenamefont {Thiessen}, \citenamefont {Bolton}, \citenamefont {Chen},
  \citenamefont {Fu}, \citenamefont {Gindulyte}, \citenamefont {Han},
  \citenamefont {He}, \citenamefont {He}, \citenamefont {Shoemaker} \emph
  {et~al.}}]{kim2016pubchem}%
  \BibitemOpen
  \bibfield  {author} {\bibinfo {author} {\bibfnamefont {S.}~\bibnamefont
  {Kim}}, \bibinfo {author} {\bibfnamefont {P.~A.}\ \bibnamefont {Thiessen}},
  \bibinfo {author} {\bibfnamefont {E.~E.}\ \bibnamefont {Bolton}}, \bibinfo
  {author} {\bibfnamefont {J.}~\bibnamefont {Chen}}, \bibinfo {author}
  {\bibfnamefont {G.}~\bibnamefont {Fu}}, \bibinfo {author} {\bibfnamefont
  {A.}~\bibnamefont {Gindulyte}}, \bibinfo {author} {\bibfnamefont
  {L.}~\bibnamefont {Han}}, \bibinfo {author} {\bibfnamefont {J.}~\bibnamefont
  {He}}, \bibinfo {author} {\bibfnamefont {S.}~\bibnamefont {He}}, \bibinfo
  {author} {\bibfnamefont {B.~A.}\ \bibnamefont {Shoemaker}},  \emph {et~al.},\
  }\href@noop {} {\bibfield  {journal} {\bibinfo  {journal} {Nucleic acids
  research}\ }\textbf {\bibinfo {volume} {44}},\ \bibinfo {pages} {D1202}
  (\bibinfo {year} {2016})}\BibitemShut {NoStop}%
\bibitem [{\citenamefont {Chen}\ and\ \citenamefont
  {Ong}(2022)}]{chen2022universal}%
  \BibitemOpen
  \bibfield  {author} {\bibinfo {author} {\bibfnamefont {C.}~\bibnamefont
  {Chen}}\ and\ \bibinfo {author} {\bibfnamefont {S.~P.}\ \bibnamefont {Ong}},\
  }\href@noop {} {\bibfield  {journal} {\bibinfo  {journal} {Nature
  Computational Science}\ }\textbf {\bibinfo {volume} {2}},\ \bibinfo {pages}
  {718} (\bibinfo {year} {2022})}\BibitemShut {NoStop}%
\bibitem [{\citenamefont {Batatia}\ \emph {et~al.}(2023)\citenamefont
  {Batatia}, \citenamefont {Benner}, \citenamefont {Chiang}, \citenamefont
  {Elena}, \citenamefont {Kov{\'a}cs}, \citenamefont {Riebesell}, \citenamefont
  {Advincula}, \citenamefont {Asta}, \citenamefont {Baldwin}, \citenamefont
  {Bernstein} \emph {et~al.}}]{batatia2023foundation}%
  \BibitemOpen
  \bibfield  {author} {\bibinfo {author} {\bibfnamefont {I.}~\bibnamefont
  {Batatia}}, \bibinfo {author} {\bibfnamefont {P.}~\bibnamefont {Benner}},
  \bibinfo {author} {\bibfnamefont {Y.}~\bibnamefont {Chiang}}, \bibinfo
  {author} {\bibfnamefont {A.~M.}\ \bibnamefont {Elena}}, \bibinfo {author}
  {\bibfnamefont {D.~P.}\ \bibnamefont {Kov{\'a}cs}}, \bibinfo {author}
  {\bibfnamefont {J.}~\bibnamefont {Riebesell}}, \bibinfo {author}
  {\bibfnamefont {X.~R.}\ \bibnamefont {Advincula}}, \bibinfo {author}
  {\bibfnamefont {M.}~\bibnamefont {Asta}}, \bibinfo {author} {\bibfnamefont
  {W.~J.}\ \bibnamefont {Baldwin}}, \bibinfo {author} {\bibfnamefont
  {N.}~\bibnamefont {Bernstein}},  \emph {et~al.},\ }\href@noop {} {\bibfield
  {journal} {\bibinfo  {journal} {arXiv preprint arXiv:2401.00096}\ } (\bibinfo
  {year} {2023})}\BibitemShut {NoStop}%
\bibitem [{\citenamefont {Hu}\ \emph {et~al.}(2023)\citenamefont {Hu},
  \citenamefont {Guo}, \citenamefont {Liu}, \citenamefont {Shi}, \citenamefont
  {Li}, \citenamefont {Hu}, \citenamefont {Bu}, \citenamefont {Luo},
  \citenamefont {He}, \citenamefont {Wang} \emph {et~al.}}]{hu2023aisnet}%
  \BibitemOpen
  \bibfield  {author} {\bibinfo {author} {\bibfnamefont {Z.}~\bibnamefont
  {Hu}}, \bibinfo {author} {\bibfnamefont {Y.}~\bibnamefont {Guo}}, \bibinfo
  {author} {\bibfnamefont {Z.}~\bibnamefont {Liu}}, \bibinfo {author}
  {\bibfnamefont {D.}~\bibnamefont {Shi}}, \bibinfo {author} {\bibfnamefont
  {Y.}~\bibnamefont {Li}}, \bibinfo {author} {\bibfnamefont {Y.}~\bibnamefont
  {Hu}}, \bibinfo {author} {\bibfnamefont {M.}~\bibnamefont {Bu}}, \bibinfo
  {author} {\bibfnamefont {K.}~\bibnamefont {Luo}}, \bibinfo {author}
  {\bibfnamefont {J.}~\bibnamefont {He}}, \bibinfo {author} {\bibfnamefont
  {C.}~\bibnamefont {Wang}},  \emph {et~al.},\ }\href@noop {} {\bibfield
  {journal} {\bibinfo  {journal} {Journal of Chemical Information and
  Modeling}\ }\textbf {\bibinfo {volume} {63}},\ \bibinfo {pages} {1756}
  (\bibinfo {year} {2023})}\BibitemShut {NoStop}%
\bibitem [{\citenamefont {Zhang}\ \emph {et~al.}(2023)\citenamefont {Zhang},
  \citenamefont {Liu}, \citenamefont {Zhang}, \citenamefont {Zhang},
  \citenamefont {Cai}, \citenamefont {Bi}, \citenamefont {Du}, \citenamefont
  {Qin}, \citenamefont {Huang}, \citenamefont {Li} \emph
  {et~al.}}]{zhang2023dpa}%
  \BibitemOpen
  \bibfield  {author} {\bibinfo {author} {\bibfnamefont {D.}~\bibnamefont
  {Zhang}}, \bibinfo {author} {\bibfnamefont {X.}~\bibnamefont {Liu}}, \bibinfo
  {author} {\bibfnamefont {X.}~\bibnamefont {Zhang}}, \bibinfo {author}
  {\bibfnamefont {C.}~\bibnamefont {Zhang}}, \bibinfo {author} {\bibfnamefont
  {C.}~\bibnamefont {Cai}}, \bibinfo {author} {\bibfnamefont {H.}~\bibnamefont
  {Bi}}, \bibinfo {author} {\bibfnamefont {Y.}~\bibnamefont {Du}}, \bibinfo
  {author} {\bibfnamefont {X.}~\bibnamefont {Qin}}, \bibinfo {author}
  {\bibfnamefont {J.}~\bibnamefont {Huang}}, \bibinfo {author} {\bibfnamefont
  {B.}~\bibnamefont {Li}},  \emph {et~al.},\ }\href {\doibase
  https://doi.org/10.48550/arXiv.2312.15492} {\bibfield  {journal} {\bibinfo
  {journal} {arXiv preprint arXiv:2312.15492}\ } (\bibinfo {year} {2023}),\
  https://doi.org/10.48550/arXiv.2312.15492}\BibitemShut {NoStop}%
\bibitem [{\citenamefont {Marks}(1990)}]{marks1990interfaces}%
  \BibitemOpen
  \bibfield  {author} {\bibinfo {author} {\bibfnamefont {T.~J.}\ \bibnamefont
  {Marks}},\ }\href {\doibase https://doi.org/10.1002/anie.199008571}
  {\bibfield  {journal} {\bibinfo  {journal} {Angewandte Chemie International
  Edition in English}\ }\textbf {\bibinfo {volume} {29}},\ \bibinfo {pages}
  {857} (\bibinfo {year} {1990})}\BibitemShut {NoStop}%
\bibitem [{\citenamefont {Chen}(1997)}]{chen1997size}%
  \BibitemOpen
  \bibfield  {author} {\bibinfo {author} {\bibfnamefont {G.}~\bibnamefont
  {Chen}},\ }\href {\doibase 10.1115/1.2824212} {\bibfield  {journal} {\bibinfo
   {journal} {Journal of Heat Transfer}\ }\textbf {\bibinfo {volume} {119}},\
  \bibinfo {pages} {220} (\bibinfo {year} {1997})}\BibitemShut {NoStop}%
\bibitem [{\citenamefont {Nan}\ \emph {et~al.}(2004)\citenamefont {Nan},
  \citenamefont {Liu}, \citenamefont {Lin},\ and\ \citenamefont
  {Li}}]{nan2004interface}%
  \BibitemOpen
  \bibfield  {author} {\bibinfo {author} {\bibfnamefont {C.-W.}\ \bibnamefont
  {Nan}}, \bibinfo {author} {\bibfnamefont {G.}~\bibnamefont {Liu}}, \bibinfo
  {author} {\bibfnamefont {Y.}~\bibnamefont {Lin}}, \ and\ \bibinfo {author}
  {\bibfnamefont {M.}~\bibnamefont {Li}},\ }\href {\doibase
  https://doi.org/10.1063/1.1808874} {\bibfield  {journal} {\bibinfo  {journal}
  {Applied Physics Letters}\ }\textbf {\bibinfo {volume} {85}},\ \bibinfo
  {pages} {3549} (\bibinfo {year} {2004})}\BibitemShut {NoStop}%
\bibitem [{\citenamefont {Burger}\ \emph {et~al.}(2016)\citenamefont {Burger},
  \citenamefont {Laachachi}, \citenamefont {Ferriol}, \citenamefont {Lutz},
  \citenamefont {Toniazzo},\ and\ \citenamefont {Ruch}}]{burger2016review}%
  \BibitemOpen
  \bibfield  {author} {\bibinfo {author} {\bibfnamefont {N.}~\bibnamefont
  {Burger}}, \bibinfo {author} {\bibfnamefont {A.}~\bibnamefont {Laachachi}},
  \bibinfo {author} {\bibfnamefont {M.}~\bibnamefont {Ferriol}}, \bibinfo
  {author} {\bibfnamefont {M.}~\bibnamefont {Lutz}}, \bibinfo {author}
  {\bibfnamefont {V.}~\bibnamefont {Toniazzo}}, \ and\ \bibinfo {author}
  {\bibfnamefont {D.}~\bibnamefont {Ruch}},\ }\href {\doibase
  https://doi.org/10.1016/j.progpolymsci.2016.05.001} {\bibfield  {journal}
  {\bibinfo  {journal} {Progress in Polymer Science}\ }\textbf {\bibinfo
  {volume} {61}},\ \bibinfo {pages} {1} (\bibinfo {year} {2016})}\BibitemShut
  {NoStop}%
\bibitem [{\citenamefont {Zhang}\ \emph {et~al.}(2019)\citenamefont {Zhang},
  \citenamefont {Hao}, \citenamefont {Nguyen}, \citenamefont {Oluwalowo},
  \citenamefont {Liu}, \citenamefont {Dessureault}, \citenamefont {Park},\ and\
  \citenamefont {Liang}}]{zhang2019carbon}%
  \BibitemOpen
  \bibfield  {author} {\bibinfo {author} {\bibfnamefont {S.}~\bibnamefont
  {Zhang}}, \bibinfo {author} {\bibfnamefont {A.}~\bibnamefont {Hao}}, \bibinfo
  {author} {\bibfnamefont {N.}~\bibnamefont {Nguyen}}, \bibinfo {author}
  {\bibfnamefont {A.}~\bibnamefont {Oluwalowo}}, \bibinfo {author}
  {\bibfnamefont {Z.}~\bibnamefont {Liu}}, \bibinfo {author} {\bibfnamefont
  {Y.}~\bibnamefont {Dessureault}}, \bibinfo {author} {\bibfnamefont {J.~G.}\
  \bibnamefont {Park}}, \ and\ \bibinfo {author} {\bibfnamefont
  {R.}~\bibnamefont {Liang}},\ }\href {\doibase
  https://doi.org/10.1016/j.carbon.2018.12.091} {\bibfield  {journal} {\bibinfo
   {journal} {Carbon}\ }\textbf {\bibinfo {volume} {144}},\ \bibinfo {pages}
  {628} (\bibinfo {year} {2019})}\BibitemShut {NoStop}%
\bibitem [{\citenamefont {Liu}\ \emph {et~al.}(2019)\citenamefont {Liu},
  \citenamefont {Wang}, \citenamefont {Shen}, \citenamefont {Qiu},
  \citenamefont {He}, \citenamefont {Zhang},\ and\ \citenamefont
  {Xie}}]{liu2019influence}%
  \BibitemOpen
  \bibfield  {author} {\bibinfo {author} {\bibfnamefont {Q.}~\bibnamefont
  {Liu}}, \bibinfo {author} {\bibfnamefont {F.}~\bibnamefont {Wang}}, \bibinfo
  {author} {\bibfnamefont {W.}~\bibnamefont {Shen}}, \bibinfo {author}
  {\bibfnamefont {X.}~\bibnamefont {Qiu}}, \bibinfo {author} {\bibfnamefont
  {Z.}~\bibnamefont {He}}, \bibinfo {author} {\bibfnamefont {Q.}~\bibnamefont
  {Zhang}}, \ and\ \bibinfo {author} {\bibfnamefont {Z.}~\bibnamefont {Xie}},\
  }\href {\doibase https://doi.org/10.1016/j.ceramint.2019.07.358} {\bibfield
  {journal} {\bibinfo  {journal} {Ceramics International}\ }\textbf {\bibinfo
  {volume} {45}},\ \bibinfo {pages} {23815} (\bibinfo {year}
  {2019})}\BibitemShut {NoStop}%
\bibitem [{\citenamefont {Bryan}\ and\ \citenamefont
  {Gamelin}(2005)}]{bryan2005doped}%
  \BibitemOpen
  \bibfield  {author} {\bibinfo {author} {\bibfnamefont {J.~D.}\ \bibnamefont
  {Bryan}}\ and\ \bibinfo {author} {\bibfnamefont {D.~R.}\ \bibnamefont
  {Gamelin}},\ }\href {\doibase https://doi.org/10.1002/0471725560.ch2}
  {\bibfield  {journal} {\bibinfo  {journal} {Progress in inorganic chemistry}\
  }\textbf {\bibinfo {volume} {54}},\ \bibinfo {pages} {47} (\bibinfo {year}
  {2005})}\BibitemShut {NoStop}%
\bibitem [{\citenamefont {Shklovskii}\ and\ \citenamefont
  {Efros}(2013)}]{shklovskii2013electronic}%
  \BibitemOpen
  \bibfield  {author} {\bibinfo {author} {\bibfnamefont {B.~I.}\ \bibnamefont
  {Shklovskii}}\ and\ \bibinfo {author} {\bibfnamefont {A.~L.}\ \bibnamefont
  {Efros}},\ }\href@noop {} {\emph {\bibinfo {title} {Electronic properties of
  doped semiconductors}}},\ Vol.~\bibinfo {volume} {45}\ (\bibinfo  {publisher}
  {Springer Science \& Business Media},\ \bibinfo {year} {2013})\BibitemShut
  {NoStop}%
\bibitem [{\citenamefont {Durand}\ \emph {et~al.}(2015)\citenamefont {Durand},
  \citenamefont {Zhang}, \citenamefont {Fowlkes}, \citenamefont {Najmaei},
  \citenamefont {Lou},\ and\ \citenamefont {Li}}]{durand2015defect}%
  \BibitemOpen
  \bibfield  {author} {\bibinfo {author} {\bibfnamefont {C.}~\bibnamefont
  {Durand}}, \bibinfo {author} {\bibfnamefont {X.}~\bibnamefont {Zhang}},
  \bibinfo {author} {\bibfnamefont {J.}~\bibnamefont {Fowlkes}}, \bibinfo
  {author} {\bibfnamefont {S.}~\bibnamefont {Najmaei}}, \bibinfo {author}
  {\bibfnamefont {J.}~\bibnamefont {Lou}}, \ and\ \bibinfo {author}
  {\bibfnamefont {A.-P.}\ \bibnamefont {Li}},\ }\href {\doibase
  https://doi.org/10.1116/1.4906331} {\bibfield  {journal} {\bibinfo  {journal}
  {Journal of Vacuum Science \& Technology B}\ }\textbf {\bibinfo {volume}
  {33}},\ \bibinfo {pages} {02B110} (\bibinfo {year} {2015})}\BibitemShut
  {NoStop}%
\bibitem [{\citenamefont {Rosenthal}(2011)}]{rosenthal2011functional}%
  \BibitemOpen
  \bibfield  {author} {\bibinfo {author} {\bibfnamefont {D.}~\bibnamefont
  {Rosenthal}},\ }\href@noop {} {\bibfield  {journal} {\bibinfo  {journal}
  {physica status solidi (a)}\ }\textbf {\bibinfo {volume} {208}},\ \bibinfo
  {pages} {1217} (\bibinfo {year} {2011})}\BibitemShut {NoStop}%
\bibitem [{\citenamefont {Huang}\ and\ \citenamefont
  {Li}(2019)}]{huang2019surface}%
  \BibitemOpen
  \bibfield  {author} {\bibinfo {author} {\bibfnamefont {W.}~\bibnamefont
  {Huang}}\ and\ \bibinfo {author} {\bibfnamefont {W.-X.}\ \bibnamefont {Li}},\
  }\href {\doibase https://doi.org/10.1039/C8CP05717F} {\bibfield  {journal}
  {\bibinfo  {journal} {Physical Chemistry Chemical Physics}\ }\textbf
  {\bibinfo {volume} {21}},\ \bibinfo {pages} {523} (\bibinfo {year}
  {2019})}\BibitemShut {NoStop}%
\bibitem [{\citenamefont {Zaera}(2021)}]{zaera2021molecular}%
  \BibitemOpen
  \bibfield  {author} {\bibinfo {author} {\bibfnamefont {F.}~\bibnamefont
  {Zaera}},\ }\href {\doibase https://doi.org/10.1016/j.ccr.2021.214179}
  {\bibfield  {journal} {\bibinfo  {journal} {Coordination Chemistry Reviews}\
  }\textbf {\bibinfo {volume} {448}},\ \bibinfo {pages} {214179} (\bibinfo
  {year} {2021})}\BibitemShut {NoStop}%
\bibitem [{\citenamefont {Biswas}\ and\ \citenamefont
  {Wu}(2005)}]{biswas2005nanoparticles}%
  \BibitemOpen
  \bibfield  {author} {\bibinfo {author} {\bibfnamefont {P.}~\bibnamefont
  {Biswas}}\ and\ \bibinfo {author} {\bibfnamefont {C.-Y.}\ \bibnamefont
  {Wu}},\ }\href {\doibase https://doi.org/10.1080/10473289.2005.10464656}
  {\bibfield  {journal} {\bibinfo  {journal} {Journal of the air \& waste
  management association}\ }\textbf {\bibinfo {volume} {55}},\ \bibinfo {pages}
  {708} (\bibinfo {year} {2005})}\BibitemShut {NoStop}%
\bibitem [{\citenamefont {Tang}\ \emph {et~al.}(2022)\citenamefont {Tang},
  \citenamefont {Simonsen}, \citenamefont {Jaganathan}, \citenamefont
  {Palot{\'a}s}, \citenamefont {Oomens}, \citenamefont {Hornek{\ae}r},\ and\
  \citenamefont {Hammer}}]{tang2022top}%
  \BibitemOpen
  \bibfield  {author} {\bibinfo {author} {\bibfnamefont {Z.}~\bibnamefont
  {Tang}}, \bibinfo {author} {\bibfnamefont {F.~D.~S.}\ \bibnamefont
  {Simonsen}}, \bibinfo {author} {\bibfnamefont {R.}~\bibnamefont
  {Jaganathan}}, \bibinfo {author} {\bibfnamefont {J.}~\bibnamefont
  {Palot{\'a}s}}, \bibinfo {author} {\bibfnamefont {J.}~\bibnamefont {Oomens}},
  \bibinfo {author} {\bibfnamefont {L.}~\bibnamefont {Hornek{\ae}r}}, \ and\
  \bibinfo {author} {\bibfnamefont {B.}~\bibnamefont {Hammer}},\ }\href@noop {}
  {\bibfield  {journal} {\bibinfo  {journal} {Astronomy \& Astrophysics}\
  }\textbf {\bibinfo {volume} {663}},\ \bibinfo {pages} {A150} (\bibinfo {year}
  {2022})}\BibitemShut {NoStop}%
\bibitem [{\citenamefont {Rasmussen}\ \emph {et~al.}(2023)\citenamefont
  {Rasmussen}, \citenamefont {Wenzel}, \citenamefont {Hornek{\ae}r},\ and\
  \citenamefont {Andersen}}]{rasmussen2023gas}%
  \BibitemOpen
  \bibfield  {author} {\bibinfo {author} {\bibfnamefont {A.~P.}\ \bibnamefont
  {Rasmussen}}, \bibinfo {author} {\bibfnamefont {G.}~\bibnamefont {Wenzel}},
  \bibinfo {author} {\bibfnamefont {L.}~\bibnamefont {Hornek{\ae}r}}, \ and\
  \bibinfo {author} {\bibfnamefont {L.~H.}\ \bibnamefont {Andersen}},\
  }\href@noop {} {\bibfield  {journal} {\bibinfo  {journal} {Astronomy \&
  Astrophysics}\ }\textbf {\bibinfo {volume} {674}},\ \bibinfo {pages} {A103}
  (\bibinfo {year} {2023})}\BibitemShut {NoStop}%
\bibitem [{\citenamefont {Poterya}\ \emph {et~al.}(2024)\citenamefont
  {Poterya}, \citenamefont {Vinkl{\'a}rek}, \citenamefont {Pysanenko},
  \citenamefont {Pluharova},\ and\ \citenamefont
  {F{\'a}rn{\'\i}k}}]{poterya2024uptake}%
  \BibitemOpen
  \bibfield  {author} {\bibinfo {author} {\bibfnamefont {V.}~\bibnamefont
  {Poterya}}, \bibinfo {author} {\bibfnamefont {I.~S.}\ \bibnamefont
  {Vinkl{\'a}rek}}, \bibinfo {author} {\bibfnamefont {A.}~\bibnamefont
  {Pysanenko}}, \bibinfo {author} {\bibfnamefont {E.}~\bibnamefont
  {Pluharova}}, \ and\ \bibinfo {author} {\bibfnamefont {M.}~\bibnamefont
  {F{\'a}rn{\'\i}k}},\ }\href {\doibase
  https://doi.org/10.1021/acsearthspacechem.3c00327} {\bibfield  {journal}
  {\bibinfo  {journal} {ACS Earth and Space Chemistry}\ }\textbf {\bibinfo
  {volume} {8}},\ \bibinfo {pages} {369} (\bibinfo {year} {2024})}\BibitemShut
  {NoStop}%
\bibitem [{\citenamefont {Ramakrishnan}\ \emph {et~al.}(2015)\citenamefont
  {Ramakrishnan}, \citenamefont {Dral}, \citenamefont {Rupp},\ and\
  \citenamefont {Von~Lilienfeld}}]{ramakrishnan2015big}%
  \BibitemOpen
  \bibfield  {author} {\bibinfo {author} {\bibfnamefont {R.}~\bibnamefont
  {Ramakrishnan}}, \bibinfo {author} {\bibfnamefont {P.~O.}\ \bibnamefont
  {Dral}}, \bibinfo {author} {\bibfnamefont {M.}~\bibnamefont {Rupp}}, \ and\
  \bibinfo {author} {\bibfnamefont {O.~A.}\ \bibnamefont {Von~Lilienfeld}},\
  }\href {\doibase https://doi.org/10.1021/acs.jctc.5b00099} {\bibfield
  {journal} {\bibinfo  {journal} {Journal of chemical theory and computation}\
  }\textbf {\bibinfo {volume} {11}},\ \bibinfo {pages} {2087} (\bibinfo {year}
  {2015})}\BibitemShut {NoStop}%
\bibitem [{\citenamefont {Christiansen}\ \emph {et~al.}(2022)\citenamefont
  {Christiansen}, \citenamefont {R{\o}nne},\ and\ \citenamefont
  {Hammer}}]{AGOX}%
  \BibitemOpen
  \bibfield  {author} {\bibinfo {author} {\bibfnamefont {M.-P.~V.}\
  \bibnamefont {Christiansen}}, \bibinfo {author} {\bibfnamefont
  {N.}~\bibnamefont {R{\o}nne}}, \ and\ \bibinfo {author} {\bibfnamefont
  {B.}~\bibnamefont {Hammer}},\ }\href@noop {} {\bibfield  {journal} {\bibinfo
  {journal} {The Journal of Chemical Physics}\ }\textbf {\bibinfo {volume}
  {157}} (\bibinfo {year} {2022})}\BibitemShut {NoStop}%
\bibitem [{\citenamefont {Larsen}\ \emph
  {et~al.}(2017{\natexlab{a}})\citenamefont {Larsen}, \citenamefont
  {Mortensen}, \citenamefont {Blomqvist}, \citenamefont {Castelli},
  \citenamefont {Christensen}, \citenamefont {Du{\l}ak}, \citenamefont {Friis},
  \citenamefont {Groves}, \citenamefont {Hammer}, \citenamefont {Hargus} \emph
  {et~al.}}]{larsen2017atomic}%
  \BibitemOpen
  \bibfield  {author} {\bibinfo {author} {\bibfnamefont {A.~H.}\ \bibnamefont
  {Larsen}}, \bibinfo {author} {\bibfnamefont {J.~J.}\ \bibnamefont
  {Mortensen}}, \bibinfo {author} {\bibfnamefont {J.}~\bibnamefont
  {Blomqvist}}, \bibinfo {author} {\bibfnamefont {I.~E.}\ \bibnamefont
  {Castelli}}, \bibinfo {author} {\bibfnamefont {R.}~\bibnamefont
  {Christensen}}, \bibinfo {author} {\bibfnamefont {M.}~\bibnamefont
  {Du{\l}ak}}, \bibinfo {author} {\bibfnamefont {J.}~\bibnamefont {Friis}},
  \bibinfo {author} {\bibfnamefont {M.~N.}\ \bibnamefont {Groves}}, \bibinfo
  {author} {\bibfnamefont {B.}~\bibnamefont {Hammer}}, \bibinfo {author}
  {\bibfnamefont {C.}~\bibnamefont {Hargus}},  \emph {et~al.},\ }\href@noop {}
  {\bibfield  {journal} {\bibinfo  {journal} {Journal of Physics: Condensed
  Matter}\ }\textbf {\bibinfo {volume} {29}},\ \bibinfo {pages} {273002}
  (\bibinfo {year} {2017}{\natexlab{a}})}\BibitemShut {NoStop}%
\bibitem [{\citenamefont {Mortensen}\ \emph {et~al.}(2024)\citenamefont
  {Mortensen}, \citenamefont {Larsen}, \citenamefont {Kuisma}, \citenamefont
  {Ivanov}, \citenamefont {Taghizadeh}, \citenamefont {Peterson}, \citenamefont
  {Haldar}, \citenamefont {Dohn}, \citenamefont {Sch{\"a}fer}, \citenamefont
  {J{\'o}nsson} \emph {et~al.}}]{GPAW}%
  \BibitemOpen
  \bibfield  {author} {\bibinfo {author} {\bibfnamefont {J.~J.}\ \bibnamefont
  {Mortensen}}, \bibinfo {author} {\bibfnamefont {A.~H.}\ \bibnamefont
  {Larsen}}, \bibinfo {author} {\bibfnamefont {M.}~\bibnamefont {Kuisma}},
  \bibinfo {author} {\bibfnamefont {A.~V.}\ \bibnamefont {Ivanov}}, \bibinfo
  {author} {\bibfnamefont {A.}~\bibnamefont {Taghizadeh}}, \bibinfo {author}
  {\bibfnamefont {A.}~\bibnamefont {Peterson}}, \bibinfo {author}
  {\bibfnamefont {A.}~\bibnamefont {Haldar}}, \bibinfo {author} {\bibfnamefont
  {A.~O.}\ \bibnamefont {Dohn}}, \bibinfo {author} {\bibfnamefont
  {C.}~\bibnamefont {Sch{\"a}fer}}, \bibinfo {author} {\bibfnamefont
  {E.~{\"O}.}\ \bibnamefont {J{\'o}nsson}},  \emph {et~al.},\ }\href@noop {}
  {\bibfield  {journal} {\bibinfo  {journal} {The Journal of Chemical Physics}\
  }\textbf {\bibinfo {volume} {160}} (\bibinfo {year} {2024})}\BibitemShut
  {NoStop}%
\bibitem [{sup()}]{supplementary}%
  \BibitemOpen
  \href@noop {} {}\bibinfo {note} {{See Supplemental Material at \textit{[URL
  will be inserted by publisher]} for further details and discussions (see also
  Refs.~\cite{perdew1996generalized,GPAW,hjorth2017ase,Escatllar2019,ronne2022atomistic,bartok2013,dscribe,dscribe2}).
  The topics covered include: CHGNet out-of-the-box performance, DFT Settings,
  Local GPR, Global Optimisation, Composition Model, Silicate Dataset Creation,
  Silver Oxide Dataset Creation}}\BibitemShut {NoStop}%
\bibitem [{\citenamefont {Slavensky}\ \emph {et~al.}(2023)\citenamefont
  {Slavensky}, \citenamefont {Christiansen},\ and\ \citenamefont
  {Hammer}}]{slavensky2023generating}%
  \BibitemOpen
  \bibfield  {author} {\bibinfo {author} {\bibfnamefont {A.~M.}\ \bibnamefont
  {Slavensky}}, \bibinfo {author} {\bibfnamefont {M.-P.~V.}\ \bibnamefont
  {Christiansen}}, \ and\ \bibinfo {author} {\bibfnamefont {B.}~\bibnamefont
  {Hammer}},\ }\href@noop {} {\bibfield  {journal} {\bibinfo  {journal} {The
  Journal of Chemical Physics}\ }\textbf {\bibinfo {volume} {159}} (\bibinfo
  {year} {2023})}\BibitemShut {NoStop}%
\bibitem [{\citenamefont {Escatllar}\ \emph {et~al.}(2019)\citenamefont
  {Escatllar}, \citenamefont {Lazaukas}, \citenamefont {Woodley},\ and\
  \citenamefont {Bromley}}]{Escatllar2019}%
  \BibitemOpen
  \bibfield  {author} {\bibinfo {author} {\bibfnamefont {A.~M.}\ \bibnamefont
  {Escatllar}}, \bibinfo {author} {\bibfnamefont {T.}~\bibnamefont {Lazaukas}},
  \bibinfo {author} {\bibfnamefont {S.~M.}\ \bibnamefont {Woodley}}, \ and\
  \bibinfo {author} {\bibfnamefont {S.~T.}\ \bibnamefont {Bromley}},\
  }\href@noop {} {\bibfield  {journal} {\bibinfo  {journal} {ACS Earth Space
  Chem.}\ }\textbf {\bibinfo {volume} {3}},\ \bibinfo {pages} {2390} (\bibinfo
  {year} {2019})}\BibitemShut {NoStop}%
\bibitem [{\citenamefont {Deng}\ \emph {et~al.}(2024)\citenamefont {Deng},
  \citenamefont {Choi}, \citenamefont {Zhong}, \citenamefont {Riebesell},
  \citenamefont {Anand}, \citenamefont {Li}, \citenamefont {Jun}, \citenamefont
  {Persson},\ and\ \citenamefont {Ceder}}]{deng2024overcoming}%
  \BibitemOpen
  \bibfield  {author} {\bibinfo {author} {\bibfnamefont {B.}~\bibnamefont
  {Deng}}, \bibinfo {author} {\bibfnamefont {Y.}~\bibnamefont {Choi}}, \bibinfo
  {author} {\bibfnamefont {P.}~\bibnamefont {Zhong}}, \bibinfo {author}
  {\bibfnamefont {J.}~\bibnamefont {Riebesell}}, \bibinfo {author}
  {\bibfnamefont {S.}~\bibnamefont {Anand}}, \bibinfo {author} {\bibfnamefont
  {Z.}~\bibnamefont {Li}}, \bibinfo {author} {\bibfnamefont {K.}~\bibnamefont
  {Jun}}, \bibinfo {author} {\bibfnamefont {K.~A.}\ \bibnamefont {Persson}}, \
  and\ \bibinfo {author} {\bibfnamefont {G.}~\bibnamefont {Ceder}},\
  }\href@noop {} {\bibfield  {journal} {\bibinfo  {journal} {arXiv preprint
  arXiv:2405.07105}\ } (\bibinfo {year} {2024})}\BibitemShut {NoStop}%
\bibitem [{\citenamefont {Focassio}\ \emph {et~al.}(2024)\citenamefont
  {Focassio}, \citenamefont {M.~Freitas},\ and\ \citenamefont
  {Schleder}}]{focassio2024performance}%
  \BibitemOpen
  \bibfield  {author} {\bibinfo {author} {\bibfnamefont {B.}~\bibnamefont
  {Focassio}}, \bibinfo {author} {\bibfnamefont {L.~P.}\ \bibnamefont
  {M.~Freitas}}, \ and\ \bibinfo {author} {\bibfnamefont {G.~R.}\ \bibnamefont
  {Schleder}},\ }\href {\doibase https://doi.org/10.1021/acsami.4c03815}
  {\bibfield  {journal} {\bibinfo  {journal} {ACS Applied Materials \&
  Interfaces}\ } (\bibinfo {year} {2024}),\
  https://doi.org/10.1021/acsami.4c03815}\BibitemShut {NoStop}%
\bibitem [{\citenamefont {K{\"a}ser}\ \emph {et~al.}(2020)\citenamefont
  {K{\"a}ser}, \citenamefont {Unke},\ and\ \citenamefont
  {Meuwly}}]{kaser2020reactive}%
  \BibitemOpen
  \bibfield  {author} {\bibinfo {author} {\bibfnamefont {S.}~\bibnamefont
  {K{\"a}ser}}, \bibinfo {author} {\bibfnamefont {O.~T.}\ \bibnamefont {Unke}},
  \ and\ \bibinfo {author} {\bibfnamefont {M.}~\bibnamefont {Meuwly}},\ }\href
  {\doibase 10.1088/1367-2630/ab81b5} {\bibfield  {journal} {\bibinfo
  {journal} {New Journal of Physics}\ }\textbf {\bibinfo {volume} {22}},\
  \bibinfo {pages} {055002} (\bibinfo {year} {2020})}\BibitemShut {NoStop}%
\bibitem [{\citenamefont {Nandi}\ \emph {et~al.}(2021)\citenamefont {Nandi},
  \citenamefont {Qu}, \citenamefont {Houston}, \citenamefont {Conte},\ and\
  \citenamefont {Bowman}}]{nandi2021delta}%
  \BibitemOpen
  \bibfield  {author} {\bibinfo {author} {\bibfnamefont {A.}~\bibnamefont
  {Nandi}}, \bibinfo {author} {\bibfnamefont {C.}~\bibnamefont {Qu}}, \bibinfo
  {author} {\bibfnamefont {P.~L.}\ \bibnamefont {Houston}}, \bibinfo {author}
  {\bibfnamefont {R.}~\bibnamefont {Conte}}, \ and\ \bibinfo {author}
  {\bibfnamefont {J.~M.}\ \bibnamefont {Bowman}},\ }\href {\doibase
  https://doi.org/10.1063/5.0038301} {\bibfield  {journal} {\bibinfo  {journal}
  {The Journal of Chemical Physics}\ }\textbf {\bibinfo {volume} {154}},\
  \bibinfo {pages} {051102} (\bibinfo {year} {2021})}\BibitemShut {NoStop}%
\bibitem [{\citenamefont {Hu}\ \emph {et~al.}(2003)\citenamefont {Hu},
  \citenamefont {Wang}, \citenamefont {Wong},\ and\ \citenamefont
  {Chen}}]{hu2003combined}%
  \BibitemOpen
  \bibfield  {author} {\bibinfo {author} {\bibfnamefont {L.}~\bibnamefont
  {Hu}}, \bibinfo {author} {\bibfnamefont {X.}~\bibnamefont {Wang}}, \bibinfo
  {author} {\bibfnamefont {L.}~\bibnamefont {Wong}}, \ and\ \bibinfo {author}
  {\bibfnamefont {G.}~\bibnamefont {Chen}},\ }\href {\doibase
  https://doi.org/10.1021/acs.jcim.3c00077} {\bibfield  {journal} {\bibinfo
  {journal} {The Journal of Chemical Physics}\ }\textbf {\bibinfo {volume}
  {119}},\ \bibinfo {pages} {11501} (\bibinfo {year} {2003})}\BibitemShut
  {NoStop}%
\bibitem [{\citenamefont {Balabin}\ and\ \citenamefont
  {Lomakina}(2009)}]{balabin2009neural}%
  \BibitemOpen
  \bibfield  {author} {\bibinfo {author} {\bibfnamefont {R.~M.}\ \bibnamefont
  {Balabin}}\ and\ \bibinfo {author} {\bibfnamefont {E.~I.}\ \bibnamefont
  {Lomakina}},\ }\href {\doibase https://doi.org/10.1063/1.3206326} {\bibfield
  {journal} {\bibinfo  {journal} {The journal of chemical physics}\ }\textbf
  {\bibinfo {volume} {131}},\ \bibinfo {pages} {074104} (\bibinfo {year}
  {2009})}\BibitemShut {NoStop}%
\bibitem [{\citenamefont {Gillan}\ \emph {et~al.}(2013)\citenamefont {Gillan},
  \citenamefont {Alf{\`e}}, \citenamefont {Bart{\'o}k},\ and\ \citenamefont
  {Cs{\'a}nyi}}]{gillan2013first}%
  \BibitemOpen
  \bibfield  {author} {\bibinfo {author} {\bibfnamefont {M.}~\bibnamefont
  {Gillan}}, \bibinfo {author} {\bibfnamefont {D.}~\bibnamefont {Alf{\`e}}},
  \bibinfo {author} {\bibfnamefont {A.}~\bibnamefont {Bart{\'o}k}}, \ and\
  \bibinfo {author} {\bibfnamefont {G.}~\bibnamefont {Cs{\'a}nyi}},\ }\href
  {\doibase https://doi.org/10.1063/1.5078687} {\bibfield  {journal} {\bibinfo
  {journal} {The Journal of chemical physics}\ }\textbf {\bibinfo {volume}
  {139}},\ \bibinfo {pages} {244504} (\bibinfo {year} {2013})}\BibitemShut
  {NoStop}%
\bibitem [{\citenamefont {Zaspel}\ \emph {et~al.}(2018)\citenamefont {Zaspel},
  \citenamefont {Huang}, \citenamefont {Harbrecht},\ and\ \citenamefont {von
  Lilienfeld}}]{zaspel2018boosting}%
  \BibitemOpen
  \bibfield  {author} {\bibinfo {author} {\bibfnamefont {P.}~\bibnamefont
  {Zaspel}}, \bibinfo {author} {\bibfnamefont {B.}~\bibnamefont {Huang}},
  \bibinfo {author} {\bibfnamefont {H.}~\bibnamefont {Harbrecht}}, \ and\
  \bibinfo {author} {\bibfnamefont {O.~A.}\ \bibnamefont {von Lilienfeld}},\
  }\href {\doibase https://doi.org/10.1021/acs.jctc.8b00832} {\bibfield
  {journal} {\bibinfo  {journal} {Journal of chemical theory and computation}\
  }\textbf {\bibinfo {volume} {15}},\ \bibinfo {pages} {1546} (\bibinfo {year}
  {2018})}\BibitemShut {NoStop}%
\bibitem [{\citenamefont {Chmiela}\ \emph {et~al.}(2018)\citenamefont
  {Chmiela}, \citenamefont {Sauceda}, \citenamefont {M{\"u}ller},\ and\
  \citenamefont {Tkatchenko}}]{chmiela2018towards}%
  \BibitemOpen
  \bibfield  {author} {\bibinfo {author} {\bibfnamefont {S.}~\bibnamefont
  {Chmiela}}, \bibinfo {author} {\bibfnamefont {H.~E.}\ \bibnamefont
  {Sauceda}}, \bibinfo {author} {\bibfnamefont {K.-R.}\ \bibnamefont
  {M{\"u}ller}}, \ and\ \bibinfo {author} {\bibfnamefont {A.}~\bibnamefont
  {Tkatchenko}},\ }\href {\doibase https://doi.org/10.1038/s41467-018-06169-2}
  {\bibfield  {journal} {\bibinfo  {journal} {Nature communications}\ }\textbf
  {\bibinfo {volume} {9}},\ \bibinfo {pages} {3887} (\bibinfo {year}
  {2018})}\BibitemShut {NoStop}%
\bibitem [{\citenamefont {Sauceda}\ \emph {et~al.}(2019)\citenamefont
  {Sauceda}, \citenamefont {Chmiela}, \citenamefont {Poltavsky}, \citenamefont
  {M{\"u}ller},\ and\ \citenamefont {Tkatchenko}}]{sauceda2019molecular}%
  \BibitemOpen
  \bibfield  {author} {\bibinfo {author} {\bibfnamefont {H.~E.}\ \bibnamefont
  {Sauceda}}, \bibinfo {author} {\bibfnamefont {S.}~\bibnamefont {Chmiela}},
  \bibinfo {author} {\bibfnamefont {I.}~\bibnamefont {Poltavsky}}, \bibinfo
  {author} {\bibfnamefont {K.-R.}\ \bibnamefont {M{\"u}ller}}, \ and\ \bibinfo
  {author} {\bibfnamefont {A.}~\bibnamefont {Tkatchenko}},\ }\href {\doibase
  https://doi.org/10.1063/1.5078687} {\bibfield  {journal} {\bibinfo  {journal}
  {The Journal of chemical physics}\ }\textbf {\bibinfo {volume} {150}},\
  \bibinfo {pages} {114102} (\bibinfo {year} {2019})}\BibitemShut {NoStop}%
\bibitem [{\citenamefont {St\"ohr}\ \emph {et~al.}(2020)\citenamefont
  {St\"ohr}, \citenamefont {Medrano~Sandonas},\ and\ \citenamefont
  {Tkatchenko}}]{stohr2020accurate}%
  \BibitemOpen
  \bibfield  {author} {\bibinfo {author} {\bibfnamefont {M.}~\bibnamefont
  {St\"ohr}}, \bibinfo {author} {\bibfnamefont {L.}~\bibnamefont
  {Medrano~Sandonas}}, \ and\ \bibinfo {author} {\bibfnamefont
  {A.}~\bibnamefont {Tkatchenko}},\ }\href {\doibase
  https://doi.org/10.1021/acs.jpclett.0c01307} {\bibfield  {journal} {\bibinfo
  {journal} {The Journal of Physical Chemistry Letters}\ }\textbf {\bibinfo
  {volume} {11}},\ \bibinfo {pages} {6835} (\bibinfo {year}
  {2020})}\BibitemShut {NoStop}%
\bibitem [{\citenamefont {Bart{\'o}k}\ \emph {et~al.}(2013)\citenamefont
  {Bart{\'o}k}, \citenamefont {Kondor},\ and\ \citenamefont
  {Cs{\'a}nyi}}]{bartok2013}%
  \BibitemOpen
  \bibfield  {author} {\bibinfo {author} {\bibfnamefont {A.~P.}\ \bibnamefont
  {Bart{\'o}k}}, \bibinfo {author} {\bibfnamefont {R.}~\bibnamefont {Kondor}},
  \ and\ \bibinfo {author} {\bibfnamefont {G.}~\bibnamefont {Cs{\'a}nyi}},\
  }\href@noop {} {\bibfield  {journal} {\bibinfo  {journal} {Physical Review
  B—Condensed Matter and Materials Physics}\ }\textbf {\bibinfo {volume}
  {87}},\ \bibinfo {pages} {184115} (\bibinfo {year} {2013})}\BibitemShut
  {NoStop}%
\bibitem [{\citenamefont {R{\o}nne}\ \emph {et~al.}(2022)\citenamefont
  {R{\o}nne}, \citenamefont {Christiansen}, \citenamefont {Slavensky},
  \citenamefont {Tang}, \citenamefont {Brix}, \citenamefont {Pedersen},
  \citenamefont {Bisbo},\ and\ \citenamefont {Hammer}}]{ronne2022atomistic}%
  \BibitemOpen
  \bibfield  {author} {\bibinfo {author} {\bibfnamefont {N.}~\bibnamefont
  {R{\o}nne}}, \bibinfo {author} {\bibfnamefont {M.-P.~V.}\ \bibnamefont
  {Christiansen}}, \bibinfo {author} {\bibfnamefont {A.~M.}\ \bibnamefont
  {Slavensky}}, \bibinfo {author} {\bibfnamefont {Z.}~\bibnamefont {Tang}},
  \bibinfo {author} {\bibfnamefont {F.}~\bibnamefont {Brix}}, \bibinfo {author}
  {\bibfnamefont {M.~E.}\ \bibnamefont {Pedersen}}, \bibinfo {author}
  {\bibfnamefont {M.~K.}\ \bibnamefont {Bisbo}}, \ and\ \bibinfo {author}
  {\bibfnamefont {B.}~\bibnamefont {Hammer}},\ }\href {\doibase
  10.1063/5.0121748} {\bibfield  {journal} {\bibinfo  {journal} {The Journal of
  Chemical Physics}\ }\textbf {\bibinfo {volume} {157}} (\bibinfo {year}
  {2022}),\ 10.1063/5.0121748}\BibitemShut {NoStop}%
\bibitem [{\citenamefont {Schnadt}\ \emph {et~al.}(2009)\citenamefont
  {Schnadt}, \citenamefont {Knudsen}, \citenamefont {Hu}, \citenamefont
  {Michaelides}, \citenamefont {Vang}, \citenamefont {Reuter}, \citenamefont
  {Li}, \citenamefont {L\ae{}gsgaard}, \citenamefont {Scheffler},\ and\
  \citenamefont {Besenbacher}}]{Ag111-O-1}%
  \BibitemOpen
  \bibfield  {author} {\bibinfo {author} {\bibfnamefont {J.}~\bibnamefont
  {Schnadt}}, \bibinfo {author} {\bibfnamefont {J.}~\bibnamefont {Knudsen}},
  \bibinfo {author} {\bibfnamefont {X.~L.}\ \bibnamefont {Hu}}, \bibinfo
  {author} {\bibfnamefont {A.}~\bibnamefont {Michaelides}}, \bibinfo {author}
  {\bibfnamefont {R.~T.}\ \bibnamefont {Vang}}, \bibinfo {author}
  {\bibfnamefont {K.}~\bibnamefont {Reuter}}, \bibinfo {author} {\bibfnamefont
  {Z.}~\bibnamefont {Li}}, \bibinfo {author} {\bibfnamefont {E.}~\bibnamefont
  {L\ae{}gsgaard}}, \bibinfo {author} {\bibfnamefont {M.}~\bibnamefont
  {Scheffler}}, \ and\ \bibinfo {author} {\bibfnamefont {F.}~\bibnamefont
  {Besenbacher}},\ }\href {\doibase 10.1103/PhysRevB.80.075424} {\bibfield
  {journal} {\bibinfo  {journal} {Phys. Rev. B}\ }\textbf {\bibinfo {volume}
  {80}},\ \bibinfo {pages} {075424} (\bibinfo {year} {2009})}\BibitemShut
  {NoStop}%
\bibitem [{\citenamefont {Derouin}\ \emph {et~al.}(2016)\citenamefont
  {Derouin}, \citenamefont {Farber}, \citenamefont {Turano}, \citenamefont
  {Iski},\ and\ \citenamefont {Killelea}}]{derouin2016thermally}%
  \BibitemOpen
  \bibfield  {author} {\bibinfo {author} {\bibfnamefont {J.}~\bibnamefont
  {Derouin}}, \bibinfo {author} {\bibfnamefont {R.~G.}\ \bibnamefont {Farber}},
  \bibinfo {author} {\bibfnamefont {M.~E.}\ \bibnamefont {Turano}}, \bibinfo
  {author} {\bibfnamefont {E.~V.}\ \bibnamefont {Iski}}, \ and\ \bibinfo
  {author} {\bibfnamefont {D.~R.}\ \bibnamefont {Killelea}},\ }\href@noop {}
  {\bibfield  {journal} {\bibinfo  {journal} {ACS Catalysis}\ }\textbf
  {\bibinfo {volume} {6}},\ \bibinfo {pages} {4640} (\bibinfo {year}
  {2016})}\BibitemShut {NoStop}%
\bibitem [{\citenamefont {Perdew}\ \emph {et~al.}(1996)\citenamefont {Perdew},
  \citenamefont {Burke},\ and\ \citenamefont
  {Ernzerhof}}]{perdew1996generalized}%
  \BibitemOpen
  \bibfield  {author} {\bibinfo {author} {\bibfnamefont {J.~P.}\ \bibnamefont
  {Perdew}}, \bibinfo {author} {\bibfnamefont {K.}~\bibnamefont {Burke}}, \
  and\ \bibinfo {author} {\bibfnamefont {M.}~\bibnamefont {Ernzerhof}},\ }\href
  {\doibase 10.1103/PhysRevLett.77.3865} {\bibfield  {journal} {\bibinfo
  {journal} {Physical review letters}\ }\textbf {\bibinfo {volume} {77}},\
  \bibinfo {pages} {3865} (\bibinfo {year} {1996})}\BibitemShut {NoStop}%
\bibitem [{\citenamefont {Larsen}\ \emph
  {et~al.}(2017{\natexlab{b}})\citenamefont {Larsen}, \citenamefont
  {Mortensen}, \citenamefont {Blomqvist}, \citenamefont {Castelli},
  \citenamefont {Christensen}, \citenamefont {Dułak}, \citenamefont {Friis},
  \citenamefont {Groves}, \citenamefont {Hammer}, \citenamefont {Hargus},
  \citenamefont {Hermes}, \citenamefont {Jennings}, \citenamefont {Jensen},
  \citenamefont {Kermode}, \citenamefont {Kitchin}, \citenamefont {Kolsbjerg},
  \citenamefont {Kubal}, \citenamefont {Kaasbjerg}, \citenamefont {Lysgaard},
  \citenamefont {Maronsson}, \citenamefont {Maxson}, \citenamefont {Olsen},
  \citenamefont {Pastewka}, \citenamefont {Peterson}, \citenamefont
  {Rostgaard}, \citenamefont {Schiøtz}, \citenamefont {Schütt}, \citenamefont
  {Strange}, \citenamefont {Thygesen}, \citenamefont {Vegge}, \citenamefont
  {Vilhelmsen}, \citenamefont {Walter}, \citenamefont {Zeng},\ and\
  \citenamefont {Jacobsen}}]{hjorth2017ase}%
  \BibitemOpen
  \bibfield  {author} {\bibinfo {author} {\bibfnamefont {A.~H.}\ \bibnamefont
  {Larsen}}, \bibinfo {author} {\bibfnamefont {J.~J.}\ \bibnamefont
  {Mortensen}}, \bibinfo {author} {\bibfnamefont {J.}~\bibnamefont
  {Blomqvist}}, \bibinfo {author} {\bibfnamefont {I.~E.}\ \bibnamefont
  {Castelli}}, \bibinfo {author} {\bibfnamefont {R.}~\bibnamefont
  {Christensen}}, \bibinfo {author} {\bibfnamefont {M.}~\bibnamefont {Dułak}},
  \bibinfo {author} {\bibfnamefont {J.}~\bibnamefont {Friis}}, \bibinfo
  {author} {\bibfnamefont {M.~N.}\ \bibnamefont {Groves}}, \bibinfo {author}
  {\bibfnamefont {B.}~\bibnamefont {Hammer}}, \bibinfo {author} {\bibfnamefont
  {C.}~\bibnamefont {Hargus}}, \bibinfo {author} {\bibfnamefont {E.~D.}\
  \bibnamefont {Hermes}}, \bibinfo {author} {\bibfnamefont {P.~C.}\
  \bibnamefont {Jennings}}, \bibinfo {author} {\bibfnamefont {P.~B.}\
  \bibnamefont {Jensen}}, \bibinfo {author} {\bibfnamefont {J.}~\bibnamefont
  {Kermode}}, \bibinfo {author} {\bibfnamefont {J.~R.}\ \bibnamefont
  {Kitchin}}, \bibinfo {author} {\bibfnamefont {E.~L.}\ \bibnamefont
  {Kolsbjerg}}, \bibinfo {author} {\bibfnamefont {J.}~\bibnamefont {Kubal}},
  \bibinfo {author} {\bibfnamefont {K.}~\bibnamefont {Kaasbjerg}}, \bibinfo
  {author} {\bibfnamefont {S.}~\bibnamefont {Lysgaard}}, \bibinfo {author}
  {\bibfnamefont {J.~B.}\ \bibnamefont {Maronsson}}, \bibinfo {author}
  {\bibfnamefont {T.}~\bibnamefont {Maxson}}, \bibinfo {author} {\bibfnamefont
  {T.}~\bibnamefont {Olsen}}, \bibinfo {author} {\bibfnamefont
  {L.}~\bibnamefont {Pastewka}}, \bibinfo {author} {\bibfnamefont
  {A.}~\bibnamefont {Peterson}}, \bibinfo {author} {\bibfnamefont
  {C.}~\bibnamefont {Rostgaard}}, \bibinfo {author} {\bibfnamefont
  {J.}~\bibnamefont {Schiøtz}}, \bibinfo {author} {\bibfnamefont
  {O.}~\bibnamefont {Schütt}}, \bibinfo {author} {\bibfnamefont
  {M.}~\bibnamefont {Strange}}, \bibinfo {author} {\bibfnamefont {K.~S.}\
  \bibnamefont {Thygesen}}, \bibinfo {author} {\bibfnamefont {T.}~\bibnamefont
  {Vegge}}, \bibinfo {author} {\bibfnamefont {L.}~\bibnamefont {Vilhelmsen}},
  \bibinfo {author} {\bibfnamefont {M.}~\bibnamefont {Walter}}, \bibinfo
  {author} {\bibfnamefont {Z.}~\bibnamefont {Zeng}}, \ and\ \bibinfo {author}
  {\bibfnamefont {K.~W.}\ \bibnamefont {Jacobsen}},\ }\href {\doibase
  10.1088/1361-648X/aa680e} {\bibfield  {journal} {\bibinfo  {journal} {Journal
  of Physics: Condensed Matter}\ }\textbf {\bibinfo {volume} {29}},\ \bibinfo
  {pages} {273002} (\bibinfo {year} {2017}{\natexlab{b}})}\BibitemShut
  {NoStop}%
\bibitem [{\citenamefont {Himanen}\ \emph {et~al.}(2020)\citenamefont
  {Himanen}, \citenamefont {J{\"a}ger}, \citenamefont {Morooka}, \citenamefont
  {Federici~Canova}, \citenamefont {Ranawat}, \citenamefont {Gao},
  \citenamefont {Rinke},\ and\ \citenamefont {Foster}}]{dscribe}%
  \BibitemOpen
  \bibfield  {author} {\bibinfo {author} {\bibfnamefont {L.}~\bibnamefont
  {Himanen}}, \bibinfo {author} {\bibfnamefont {M.~O.~J.}\ \bibnamefont
  {J{\"a}ger}}, \bibinfo {author} {\bibfnamefont {E.~V.}\ \bibnamefont
  {Morooka}}, \bibinfo {author} {\bibfnamefont {F.}~\bibnamefont
  {Federici~Canova}}, \bibinfo {author} {\bibfnamefont {Y.~S.}\ \bibnamefont
  {Ranawat}}, \bibinfo {author} {\bibfnamefont {D.~Z.}\ \bibnamefont {Gao}},
  \bibinfo {author} {\bibfnamefont {P.}~\bibnamefont {Rinke}}, \ and\ \bibinfo
  {author} {\bibfnamefont {A.~S.}\ \bibnamefont {Foster}},\ }\href {\doibase
  10.1016/j.cpc.2019.106949} {\bibfield  {journal} {\bibinfo  {journal}
  {Computer Physics Communications}\ }\textbf {\bibinfo {volume} {247}},\
  \bibinfo {pages} {106949} (\bibinfo {year} {2020})}\BibitemShut {NoStop}%
\bibitem [{\citenamefont {Laakso}\ \emph {et~al.}(2023)\citenamefont {Laakso},
  \citenamefont {Himanen}, \citenamefont {Homm}, \citenamefont {Morooka},
  \citenamefont {J{\"a}ger}, \citenamefont {Todorovi{\'c}},\ and\ \citenamefont
  {Rinke}}]{dscribe2}%
  \BibitemOpen
  \bibfield  {author} {\bibinfo {author} {\bibfnamefont {J.}~\bibnamefont
  {Laakso}}, \bibinfo {author} {\bibfnamefont {L.}~\bibnamefont {Himanen}},
  \bibinfo {author} {\bibfnamefont {H.}~\bibnamefont {Homm}}, \bibinfo {author}
  {\bibfnamefont {E.~V.}\ \bibnamefont {Morooka}}, \bibinfo {author}
  {\bibfnamefont {M.~O.}\ \bibnamefont {J{\"a}ger}}, \bibinfo {author}
  {\bibfnamefont {M.}~\bibnamefont {Todorovi{\'c}}}, \ and\ \bibinfo {author}
  {\bibfnamefont {P.}~\bibnamefont {Rinke}},\ }\href@noop {} {\bibfield
  {journal} {\bibinfo  {journal} {The Journal of Chemical Physics}\ }\textbf
  {\bibinfo {volume} {158}} (\bibinfo {year} {2023})}\BibitemShut {NoStop}%
\end{thebibliography}%


\begin{thebibliography}{9}%
\makeatletter
\providecommand \@ifxundefined [1]{%
 \@ifx{#1\undefined}
}%
\providecommand \@ifnum [1]{%
 \ifnum #1\expandafter \@firstoftwo
 \else \expandafter \@secondoftwo
 \fi
}%
\providecommand \@ifx [1]{%
 \ifx #1\expandafter \@firstoftwo
 \else \expandafter \@secondoftwo
 \fi
}%
\providecommand \natexlab [1]{#1}%
\providecommand \enquote  [1]{``#1''}%
\providecommand \bibnamefont  [1]{#1}%
\providecommand \bibfnamefont [1]{#1}%
\providecommand \citenamefont [1]{#1}%
\providecommand \href@noop [0]{\@secondoftwo}%
\providecommand \href [0]{\begingroup \@sanitize@url \@href}%
\providecommand \@href[1]{\@@startlink{#1}\@@href}%
\providecommand \@@href[1]{\endgroup#1\@@endlink}%
\providecommand \@sanitize@url [0]{\catcode `\\12\catcode `\$12\catcode
  `\&12\catcode `\#12\catcode `\^12\catcode `\_12\catcode `\%12\relax}%
\providecommand \@@startlink[1]{}%
\providecommand \@@endlink[0]{}%
\providecommand \url  [0]{\begingroup\@sanitize@url \@url }%
\providecommand \@url [1]{\endgroup\@href {#1}{\urlprefix }}%
\providecommand \urlprefix  [0]{URL }%
\providecommand \Eprint [0]{\href }%
\providecommand \doibase [0]{http://dx.doi.org/}%
\providecommand \selectlanguage [0]{\@gobble}%
\providecommand \bibinfo  [0]{\@secondoftwo}%
\providecommand \bibfield  [0]{\@secondoftwo}%
\providecommand \translation [1]{[#1]}%
\providecommand \BibitemOpen [0]{}%
\providecommand \bibitemStop [0]{}%
\providecommand \bibitemNoStop [0]{.\EOS\space}%
\providecommand \EOS [0]{\spacefactor3000\relax}%
\providecommand \BibitemShut  [1]{\csname bibitem#1\endcsname}%
\let\auto@bib@innerbib\@empty
\bibitem [{\citenamefont {Escatllar}\ \emph {et~al.}(2019)\citenamefont
  {Escatllar}, \citenamefont {Lazaukas}, \citenamefont {Woodley},\ and\
  \citenamefont {Bromley}}]{Escatllar2019}%
  \BibitemOpen
  \bibfield  {author} {\bibinfo {author} {\bibfnamefont {A.~M.}\ \bibnamefont
  {Escatllar}}, \bibinfo {author} {\bibfnamefont {T.}~\bibnamefont {Lazaukas}},
  \bibinfo {author} {\bibfnamefont {S.~M.}\ \bibnamefont {Woodley}}, \ and\
  \bibinfo {author} {\bibfnamefont {S.~T.}\ \bibnamefont {Bromley}},\
  }\href@noop {} {\bibfield  {journal} {\bibinfo  {journal} {ACS Earth Space
  Chem.}\ }\textbf {\bibinfo {volume} {3}},\ \bibinfo {pages} {2390} (\bibinfo
  {year} {2019})}\BibitemShut {NoStop}%
\bibitem [{\citenamefont {Mortensen}\ \emph {et~al.}(2024)\citenamefont
  {Mortensen}, \citenamefont {Larsen}, \citenamefont {Kuisma}, \citenamefont
  {Ivanov}, \citenamefont {Taghizadeh}, \citenamefont {Peterson}, \citenamefont
  {Haldar}, \citenamefont {Dohn}, \citenamefont {Sch{\"a}fer}, \citenamefont
  {J{\'o}nsson} \emph {et~al.}}]{GPAW}%
  \BibitemOpen
  \bibfield  {author} {\bibinfo {author} {\bibfnamefont {J.~J.}\ \bibnamefont
  {Mortensen}}, \bibinfo {author} {\bibfnamefont {A.~H.}\ \bibnamefont
  {Larsen}}, \bibinfo {author} {\bibfnamefont {M.}~\bibnamefont {Kuisma}},
  \bibinfo {author} {\bibfnamefont {A.~V.}\ \bibnamefont {Ivanov}}, \bibinfo
  {author} {\bibfnamefont {A.}~\bibnamefont {Taghizadeh}}, \bibinfo {author}
  {\bibfnamefont {A.}~\bibnamefont {Peterson}}, \bibinfo {author}
  {\bibfnamefont {A.}~\bibnamefont {Haldar}}, \bibinfo {author} {\bibfnamefont
  {A.~O.}\ \bibnamefont {Dohn}}, \bibinfo {author} {\bibfnamefont
  {C.}~\bibnamefont {Sch{\"a}fer}}, \bibinfo {author} {\bibfnamefont
  {E.~{\"O}.}\ \bibnamefont {J{\'o}nsson}},  \emph {et~al.},\ }\href@noop {}
  {\bibfield  {journal} {\bibinfo  {journal} {The Journal of Chemical Physics}\
  }\textbf {\bibinfo {volume} {160}} (\bibinfo {year} {2024})}\BibitemShut
  {NoStop}%
\bibitem [{\citenamefont {Larsen}\ \emph {et~al.}(2017)\citenamefont {Larsen},
  \citenamefont {Mortensen}, \citenamefont {Blomqvist}, \citenamefont
  {Castelli}, \citenamefont {Christensen}, \citenamefont {Dułak},
  \citenamefont {Friis}, \citenamefont {Groves}, \citenamefont {Hammer},
  \citenamefont {Hargus}, \citenamefont {Hermes}, \citenamefont {Jennings},
  \citenamefont {Jensen}, \citenamefont {Kermode}, \citenamefont {Kitchin},
  \citenamefont {Kolsbjerg}, \citenamefont {Kubal}, \citenamefont {Kaasbjerg},
  \citenamefont {Lysgaard}, \citenamefont {Maronsson}, \citenamefont {Maxson},
  \citenamefont {Olsen}, \citenamefont {Pastewka}, \citenamefont {Peterson},
  \citenamefont {Rostgaard}, \citenamefont {Schiøtz}, \citenamefont {Schütt},
  \citenamefont {Strange}, \citenamefont {Thygesen}, \citenamefont {Vegge},
  \citenamefont {Vilhelmsen}, \citenamefont {Walter}, \citenamefont {Zeng},\
  and\ \citenamefont {Jacobsen}}]{hjorth2017ase}%
  \BibitemOpen
  \bibfield  {author} {\bibinfo {author} {\bibfnamefont {A.~H.}\ \bibnamefont
  {Larsen}}, \bibinfo {author} {\bibfnamefont {J.~J.}\ \bibnamefont
  {Mortensen}}, \bibinfo {author} {\bibfnamefont {J.}~\bibnamefont
  {Blomqvist}}, \bibinfo {author} {\bibfnamefont {I.~E.}\ \bibnamefont
  {Castelli}}, \bibinfo {author} {\bibfnamefont {R.}~\bibnamefont
  {Christensen}}, \bibinfo {author} {\bibfnamefont {M.}~\bibnamefont {Dułak}},
  \bibinfo {author} {\bibfnamefont {J.}~\bibnamefont {Friis}}, \bibinfo
  {author} {\bibfnamefont {M.~N.}\ \bibnamefont {Groves}}, \bibinfo {author}
  {\bibfnamefont {B.}~\bibnamefont {Hammer}}, \bibinfo {author} {\bibfnamefont
  {C.}~\bibnamefont {Hargus}}, \bibinfo {author} {\bibfnamefont {E.~D.}\
  \bibnamefont {Hermes}}, \bibinfo {author} {\bibfnamefont {P.~C.}\
  \bibnamefont {Jennings}}, \bibinfo {author} {\bibfnamefont {P.~B.}\
  \bibnamefont {Jensen}}, \bibinfo {author} {\bibfnamefont {J.}~\bibnamefont
  {Kermode}}, \bibinfo {author} {\bibfnamefont {J.~R.}\ \bibnamefont
  {Kitchin}}, \bibinfo {author} {\bibfnamefont {E.~L.}\ \bibnamefont
  {Kolsbjerg}}, \bibinfo {author} {\bibfnamefont {J.}~\bibnamefont {Kubal}},
  \bibinfo {author} {\bibfnamefont {K.}~\bibnamefont {Kaasbjerg}}, \bibinfo
  {author} {\bibfnamefont {S.}~\bibnamefont {Lysgaard}}, \bibinfo {author}
  {\bibfnamefont {J.~B.}\ \bibnamefont {Maronsson}}, \bibinfo {author}
  {\bibfnamefont {T.}~\bibnamefont {Maxson}}, \bibinfo {author} {\bibfnamefont
  {T.}~\bibnamefont {Olsen}}, \bibinfo {author} {\bibfnamefont
  {L.}~\bibnamefont {Pastewka}}, \bibinfo {author} {\bibfnamefont
  {A.}~\bibnamefont {Peterson}}, \bibinfo {author} {\bibfnamefont
  {C.}~\bibnamefont {Rostgaard}}, \bibinfo {author} {\bibfnamefont
  {J.}~\bibnamefont {Schiøtz}}, \bibinfo {author} {\bibfnamefont
  {O.}~\bibnamefont {Schütt}}, \bibinfo {author} {\bibfnamefont
  {M.}~\bibnamefont {Strange}}, \bibinfo {author} {\bibfnamefont {K.~S.}\
  \bibnamefont {Thygesen}}, \bibinfo {author} {\bibfnamefont {T.}~\bibnamefont
  {Vegge}}, \bibinfo {author} {\bibfnamefont {L.}~\bibnamefont {Vilhelmsen}},
  \bibinfo {author} {\bibfnamefont {M.}~\bibnamefont {Walter}}, \bibinfo
  {author} {\bibfnamefont {Z.}~\bibnamefont {Zeng}}, \ and\ \bibinfo {author}
  {\bibfnamefont {K.~W.}\ \bibnamefont {Jacobsen}},\ }\href {\doibase
  10.1088/1361-648X/aa680e} {\bibfield  {journal} {\bibinfo  {journal} {Journal
  of Physics: Condensed Matter}\ }\textbf {\bibinfo {volume} {29}},\ \bibinfo
  {pages} {273002} (\bibinfo {year} {2017})}\BibitemShut {NoStop}%
\bibitem [{\citenamefont {Perdew}\ \emph {et~al.}(1996)\citenamefont {Perdew},
  \citenamefont {Burke},\ and\ \citenamefont
  {Ernzerhof}}]{perdew1996generalized}%
  \BibitemOpen
  \bibfield  {author} {\bibinfo {author} {\bibfnamefont {J.~P.}\ \bibnamefont
  {Perdew}}, \bibinfo {author} {\bibfnamefont {K.}~\bibnamefont {Burke}}, \
  and\ \bibinfo {author} {\bibfnamefont {M.}~\bibnamefont {Ernzerhof}},\ }\href
  {\doibase 10.1103/PhysRevLett.77.3865} {\bibfield  {journal} {\bibinfo
  {journal} {Physical review letters}\ }\textbf {\bibinfo {volume} {77}},\
  \bibinfo {pages} {3865} (\bibinfo {year} {1996})}\BibitemShut {NoStop}%
\bibitem [{\citenamefont {R{\o}nne}\ \emph {et~al.}(2022)\citenamefont
  {R{\o}nne}, \citenamefont {Christiansen}, \citenamefont {Slavensky},
  \citenamefont {Tang}, \citenamefont {Brix}, \citenamefont {Pedersen},
  \citenamefont {Bisbo},\ and\ \citenamefont {Hammer}}]{ronne2022atomistic}%
  \BibitemOpen
  \bibfield  {author} {\bibinfo {author} {\bibfnamefont {N.}~\bibnamefont
  {R{\o}nne}}, \bibinfo {author} {\bibfnamefont {M.-P.~V.}\ \bibnamefont
  {Christiansen}}, \bibinfo {author} {\bibfnamefont {A.~M.}\ \bibnamefont
  {Slavensky}}, \bibinfo {author} {\bibfnamefont {Z.}~\bibnamefont {Tang}},
  \bibinfo {author} {\bibfnamefont {F.}~\bibnamefont {Brix}}, \bibinfo {author}
  {\bibfnamefont {M.~E.}\ \bibnamefont {Pedersen}}, \bibinfo {author}
  {\bibfnamefont {M.~K.}\ \bibnamefont {Bisbo}}, \ and\ \bibinfo {author}
  {\bibfnamefont {B.}~\bibnamefont {Hammer}},\ }\href {\doibase
  10.1063/5.0121748} {\bibfield  {journal} {\bibinfo  {journal} {The Journal of
  Chemical Physics}\ }\textbf {\bibinfo {volume} {157}} (\bibinfo {year}
  {2022}),\ 10.1063/5.0121748}\BibitemShut {NoStop}%
\bibitem [{\citenamefont {Bart{\'o}k}\ \emph {et~al.}(2013)\citenamefont
  {Bart{\'o}k}, \citenamefont {Kondor},\ and\ \citenamefont
  {Cs{\'a}nyi}}]{bartok2013}%
  \BibitemOpen
  \bibfield  {author} {\bibinfo {author} {\bibfnamefont {A.~P.}\ \bibnamefont
  {Bart{\'o}k}}, \bibinfo {author} {\bibfnamefont {R.}~\bibnamefont {Kondor}},
  \ and\ \bibinfo {author} {\bibfnamefont {G.}~\bibnamefont {Cs{\'a}nyi}},\
  }\href@noop {} {\bibfield  {journal} {\bibinfo  {journal} {Physical Review
  B—Condensed Matter and Materials Physics}\ }\textbf {\bibinfo {volume}
  {87}},\ \bibinfo {pages} {184115} (\bibinfo {year} {2013})}\BibitemShut
  {NoStop}%
\bibitem [{\citenamefont {Himanen}\ \emph {et~al.}(2020)\citenamefont
  {Himanen}, \citenamefont {J{\"a}ger}, \citenamefont {Morooka}, \citenamefont
  {Federici~Canova}, \citenamefont {Ranawat}, \citenamefont {Gao},
  \citenamefont {Rinke},\ and\ \citenamefont {Foster}}]{dscribe}%
  \BibitemOpen
  \bibfield  {author} {\bibinfo {author} {\bibfnamefont {L.}~\bibnamefont
  {Himanen}}, \bibinfo {author} {\bibfnamefont {M.~O.~J.}\ \bibnamefont
  {J{\"a}ger}}, \bibinfo {author} {\bibfnamefont {E.~V.}\ \bibnamefont
  {Morooka}}, \bibinfo {author} {\bibfnamefont {F.}~\bibnamefont
  {Federici~Canova}}, \bibinfo {author} {\bibfnamefont {Y.~S.}\ \bibnamefont
  {Ranawat}}, \bibinfo {author} {\bibfnamefont {D.~Z.}\ \bibnamefont {Gao}},
  \bibinfo {author} {\bibfnamefont {P.}~\bibnamefont {Rinke}}, \ and\ \bibinfo
  {author} {\bibfnamefont {A.~S.}\ \bibnamefont {Foster}},\ }\href {\doibase
  10.1016/j.cpc.2019.106949} {\bibfield  {journal} {\bibinfo  {journal}
  {Computer Physics Communications}\ }\textbf {\bibinfo {volume} {247}},\
  \bibinfo {pages} {106949} (\bibinfo {year} {2020})}\BibitemShut {NoStop}%
\bibitem [{\citenamefont {Laakso}\ \emph {et~al.}(2023)\citenamefont {Laakso},
  \citenamefont {Himanen}, \citenamefont {Homm}, \citenamefont {Morooka},
  \citenamefont {J{\"a}ger}, \citenamefont {Todorovi{\'c}},\ and\ \citenamefont
  {Rinke}}]{dscribe2}%
  \BibitemOpen
  \bibfield  {author} {\bibinfo {author} {\bibfnamefont {J.}~\bibnamefont
  {Laakso}}, \bibinfo {author} {\bibfnamefont {L.}~\bibnamefont {Himanen}},
  \bibinfo {author} {\bibfnamefont {H.}~\bibnamefont {Homm}}, \bibinfo {author}
  {\bibfnamefont {E.~V.}\ \bibnamefont {Morooka}}, \bibinfo {author}
  {\bibfnamefont {M.~O.}\ \bibnamefont {J{\"a}ger}}, \bibinfo {author}
  {\bibfnamefont {M.}~\bibnamefont {Todorovi{\'c}}}, \ and\ \bibinfo {author}
  {\bibfnamefont {P.}~\bibnamefont {Rinke}},\ }\href@noop {} {\bibfield
  {journal} {\bibinfo  {journal} {The Journal of Chemical Physics}\ }\textbf
  {\bibinfo {volume} {158}} (\bibinfo {year} {2023})}\BibitemShut {NoStop}%
\bibitem [{\citenamefont {Christiansen}\ \emph {et~al.}(2022)\citenamefont
  {Christiansen}, \citenamefont {R{\o}nne},\ and\ \citenamefont
  {Hammer}}]{AGOX}%
  \BibitemOpen
  \bibfield  {author} {\bibinfo {author} {\bibfnamefont {M.-P.~V.}\
  \bibnamefont {Christiansen}}, \bibinfo {author} {\bibfnamefont
  {N.}~\bibnamefont {R{\o}nne}}, \ and\ \bibinfo {author} {\bibfnamefont
  {B.}~\bibnamefont {Hammer}},\ }\href@noop {} {\bibfield  {journal} {\bibinfo
  {journal} {The Journal of Chemical Physics}\ }\textbf {\bibinfo {volume}
  {157}} (\bibinfo {year} {2022})}\BibitemShut {NoStop}%
\end{thebibliography}%
\bibliographystyle{apsrev4-1}

\end{document}


\title{Supplementary Information}
\author{Joe Pitfield}\email{joepitfield@gmail.com}

\author{Florian Brix}
\author{Zeyuan Tang}
\author{Andreas Møller Slavensky}
\author{Nikolaj Rønne}
\author{Mads-Peter Verner Christiansen}
\author{Bjørk Hammer}
\email{hammer@phys.au.dk}

\maketitle

\section{Other examples of CHGNet success OOTB} 

\begin{figure}[h!]
  \centering
  \includegraphics[width=\linewidth]{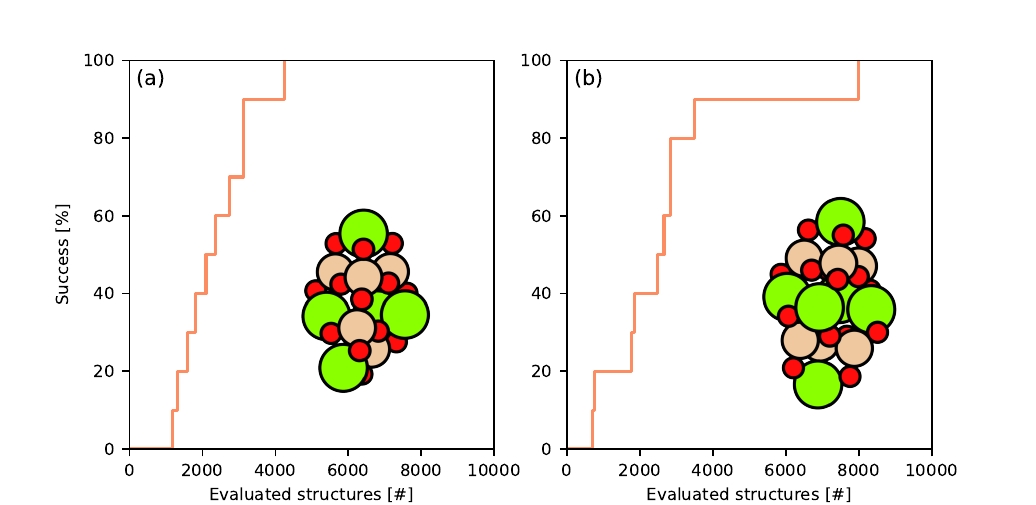}
  \caption{Success curves for finding the global energy minimum
    structures of pyroxene with \textbf{(a)} five and \textbf{(b)} six
    formula units.
    The target potential is CHGNet.
  }
  \label{fig:success_p5_p6}
\end{figure}

Success curves for global optimization of pyroxene with five and six formula units, using
CHGNet as the target potential, are shown in figure S\ref{fig:success_p5_p6}. The best structures found in CHGNet indeed
coincide with those found previously when using DFT as the target potential\cite{Escatllar2019}.
This testifies to the extrapolative power of the pretrained CHGNet.

\newpage
In Fig.\ \figref[]{fig:cu111_chg}, the results from a structural
optimization for $(2\sqrt{2}\times\sqrt{2})\mathrm{R}45^\circ$
O-Cu(100) are considered. The search was done with DFT as the target
potential, and subsequently the ideal and some defected structures
were analysed with the pretrained CHGNet. The strong correlation
demonstrates the predictive power of pretrained CHGNet for this
system, where no augmentation is required.

\begin{figure}[h!]
    \centering
    \includegraphics[width=0.5\linewidth]{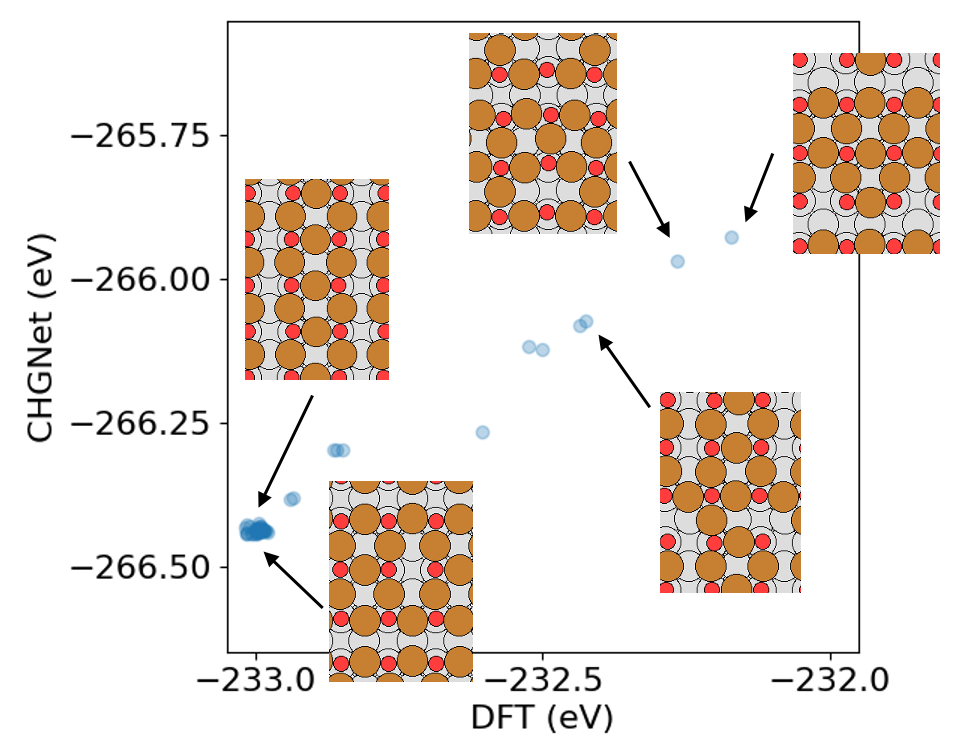}
    \caption{Parity plot of CHGNet versus DFT energies for various
      ideal and defected $(2\sqrt{2}\times\sqrt{2})\mathrm{R}45^\circ$
      O-Cu(100) surface structures.
    }
    \label{fig:cu111_chg}
\end{figure}
\section{DFT settings} 

For all first-principles calculations performed throughout this work,
we employed Density Functional Theory (DFT) as implemented in the GPAW
package \cite{GPAW} and handled via the Atomic Simulation Environment
package\cite{hjorth2017ase}. GPAW was used in the plane-wave mode with an energy
cutoff of 500 eV. The XC-correlation functional adopted was the
PBE\cite{perdew1996generalized}. The lattice constant of {\sc fcc}-Ag was 4.15 {\AA}. For
modelling the Ag(111) surfaces, slabs of three (111) layers, with the
two bottom layers fixed, were introduced. Forces were relaxed until
being smaller than 0.025 eV/{\AA}. For the Ag(111)-$c(4\times 8)$,
Ag(111)-$c(3\times 5\sqrt{3})$, and Ag(111)-$p(4\times 5\sqrt{3})$
surface unit cells, planar $\mathbf{k}$-point sets of $(6\times 6)$,
$(4\times 4)$, and $(4\times 2)$ were used.

\section{Local GPR} \label{sec:sgpr} 

In prior work, we introduced a sparse local GPR model
\cite{ronne2022atomistic}. We use a radial basis function kernel
to evaluate covariance between local environments described through
the SOAP representation\cite{bartok2013}, as implemented in DScribe~\cite{dscribe,dscribe2}, given as
\begin{equation}
	k(\mathbf{x}_i, \mathbf{x}_j) = A
        \exp\left(-\frac{|\mathbf{x}_i-\mathbf{x}_j|^2}{2l^2} \right).
	\label{eq:kernel}
\end{equation}
$A$ and $l$ are amplitude and length-scale hyperparameters.
Predictions of local residual energies from the GPR is given by
\begin{equation}
\Delta E^{\mathrm{GPR}}(\mathbf{x}_i) = \mathbf{k}_m(\mathbf{x}_i)C
\left[\mathbf{E^{\mathrm{DFT}}}-\mathbf{E^{\mathrm{CHGNet}}} \right],
\label{eq:sparse_lgpr_energy}
\end{equation}
where $\mathbf{x}_i$ is the local atomic representation of atom $i$,
$\mathbf{k}_m(\mathbf{x}_i)$ is the kernel between $\mathbf{x}_i$ and the
inducing points $X_m$, the $C$ matrix is the sparse version of the inverted covariance matrix
and $\mathbf{E^{\mathrm{DFT}}}-\mathbf{E^{\mathrm{CHGNet}}}$ is the
training dataset residual energies. The $C$ matrix is defined through the
training dataset representations $X_n$ and a set of inducing points
$X_m$ as
\begin{equation}
	C = \left[K_{mm} + (LK_{nm})^T \Sigma_{nn}^{-1}(LK_{nm})\right]^{-1}(LK_{nm})^T\Sigma^{-1},
\end{equation}
where $\Sigma_{nn}$ is a diagonal matrix of local noise,
$K_{mm} = k(X_m, X_m)$,
$K_{nm} = K_{mn}^{T} = k(X_n, X_m)$ and $L$ the local energy correspondence
matrix. The $C\left[\mathbf{E^{\mathrm{DFT}}}-\mathbf{E^{\mathrm{CHGNet}}} \right]$ 
vector is pre-calculated during training.

An identical formulation can be constructed for implementing 
forces into the training data,  
where including gradients of the energy landscape invariably 
increases the accuracy of the model at the cost of decreasing the 
quantity of structures with which the GPR remains performative. 
\newpage

\section{Global optimization} 

Searches were performed using parallel tempering as described in Ref.\
\onlinecite{AGOX}. 10 walkers were introduced with temperatures
distributed exponentially from $k_BT=0.02$ eV to $2.0$ eV. In every
step, the structure of each walker was rattled randomly and relaxed in
a local-descriptor sparse GPR model-based interatomic potential. The new structures were accepted
according to the Metropolis Monte Carlo criterion. Rattle amplitudes
were adjusted to achieve a 50\% acceptance ratios. Swaps of structures
between walkers were attempted for every 10 episodes. DFT calculations
of the structures of the five lowest-temperature walkers were done for
every 50 episodes, and the GPR model was updated. Runs were performed
until the best structure, the putative GM structure, had been found
multiple times for each stoichiometry.

The local descriptor used in the silicate cluster searches was SOAP-based with
$r_{cut}=7$ {\AA}, $n_{max}=3$, and $l_{max}=2$. The kernel was a radial basis
function kernel (RBF) with length-scale = $l=20$ and prefactor = $A=1$.

The local descriptor and kernel used in the silver oxide film searches were also
SOAP-based and RBF-type, however with $r_{cut}=7$ {\AA}, $n_{max}=3=$,
$l_{max}=2=$, length-scale $l=100$, and prefactor $A=1$.

\newpage
\section{Compositional Model} 
The CHGNet architecture includes a composition model that represents elemental one-body terms but crucially also average interaction energies for each element. This is a linear model
$$
E_{comp}(S) = \sum_z n_z \epsilon_z,
$$
where $n_z$ is the number of atoms of species $z$ in structure $S$ and $\epsilon_z$ is the learned average elemental energy for an atom of type $z$.

When training on DFT calculations produced with different settings
than those of the original MPtrj training set, controlling this model
is crucial. This translates into finding a new composition model,
$E_{comp}$, which provides the best predictions when replacing the
given composition model, $E^0_{comp}$, contained in the original
CHGNet model, $E^0_{CHG}$. I.e.\ we seek the $E_{comp}$ that best fulfills:
$$
E_{DFT}(S) \approx E^0_{CHG}(S) - E^0_{comp}(S) + E_{comp}(S).
$$
To refit this composition model on a new dataset we therefore do so with labels given by:
$$
E_C^\ast(S) = E_{DFT}(S) - [E^0_{CHG}(S)-E^0_{comp}(S)],
$$
where the term in the brackets is the CHGNet prediction of the difference from the average interaction energy. We fit the composition model by minimizing the least squares loss wrt. $\epsilon_z$:
$$
L = \frac{1}{N}\sum_S ||E_C^\ast(S) - E_{comp}(S)||^2,
$$
where $N$ is the total number of structures in the training set. We use 
the labels $E^\ast_C$ when refitting the composition model because if $E_{DFT}$ was used, the new refitted model would essentially double count some parts of the interaction energy, as it would be accounted for both by the network and the composition model.

\newpage

\section{Silicate Dataset Creation} 

The dataset for PYR-4 used in Fig.\ 2 was created in the follow way:

\begin{itemize}
\item
Thirty independent search campaigns, following the global optimistiton 
methodology outlined in Section S{\sc IV}, were conducted each producing about
100 structures.
\item
All structures were combined resulting in a total DFT dataset 
consisting of 3098 structures.
\item
DFT evaluations were performed after every 1000 episodes.
\end{itemize}

\section{Silver oxide dataset creation} 

\begin{figure}
    \centering
    \includegraphics[width=\linewidth]{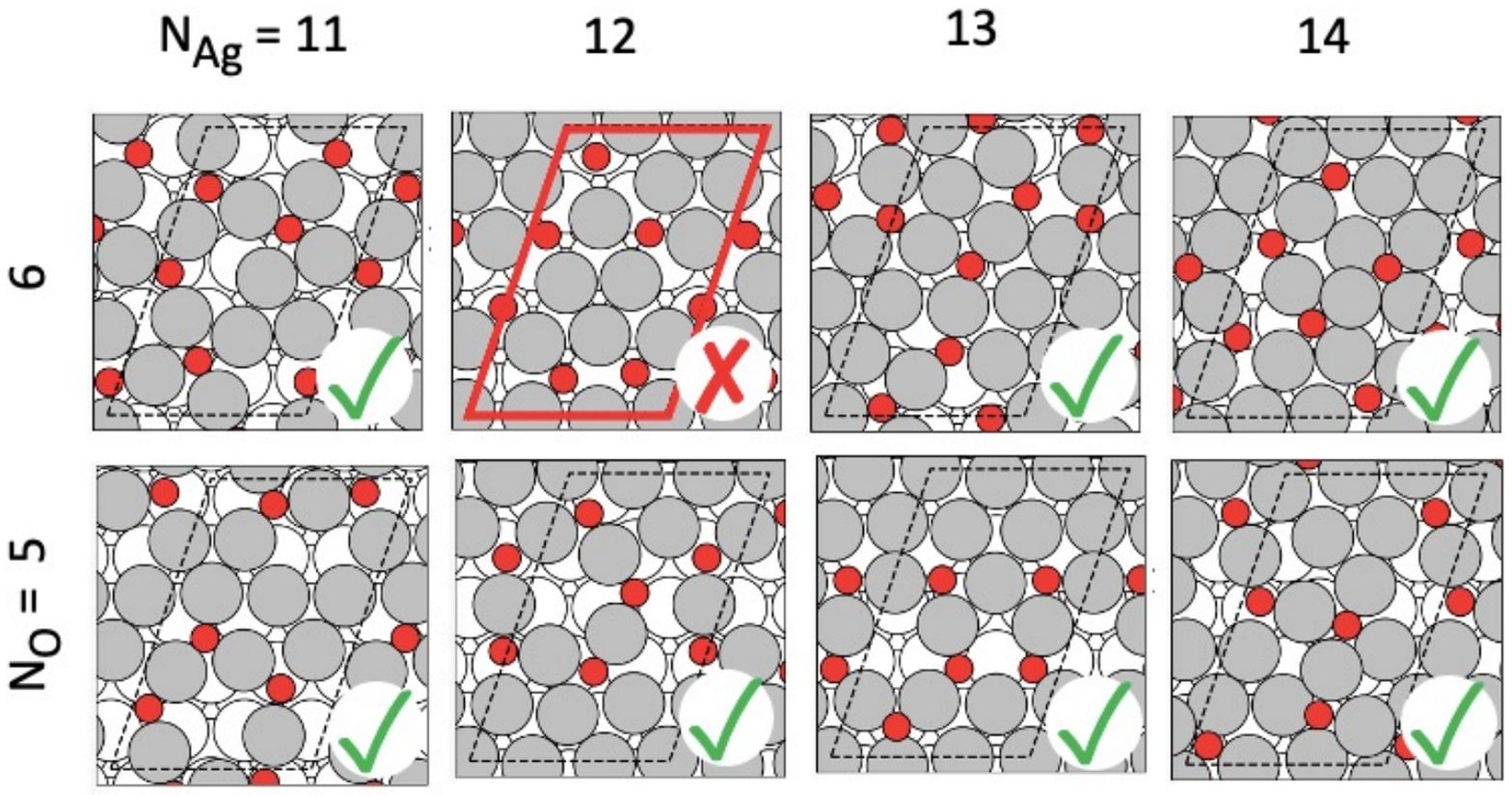}
    \caption{Images of the most energetically favourable structures 
    found by a set of independent active learning runs over a range
    of stoichiometries for the O-Ag(111) surface reconstruction. 
    The offset structure indicates that that calculation did not 
    contribute to the final dataset employed in the main body of the 
    letter, but did indeed correctly identify the dft ground state 
    structure.}
    \label{fig:ag111-c3}
\end{figure}

The dataset for silver oxide thin films used in Fig.\ 3 was created in the follow way:

\begin{itemize}
\item Thin films of stoichiometry Ag$_5$O$_{11}$, Ag$_5$O$_{12}$,
  Ag$_5$O$_{13}$, Ag$_5$O$_{14}$, Ag$_6$O$_{11}$, Ag$_6$O$_{13}$, or
  Ag$_6$O$_{14}$ were considered while films of Ag$_6$O$_{12}$ were
  neglected.
\item
For each stoichometry ten independent search campaigns were conducted
using the parallel tempering search strategy outlined in Section S{\sc IV}. The GM
structures found are depicted in Fig.\ S\ref{fig:ag111-c3}
\item
The search results were combined into about 80 structures per
stoichiometry, and a total 640 structures for all stoichiometries.
\end{itemize}

\bibliography{biblio}
\bibliographystyle{apsrev4-1}